\documentclass{pasj01}

\begin{document}
\SetRunningHead{Makishima et al.}{Precession in 1E 1547.0$-$5408}


\title{
Evidence for a 36 ks Phase Modulation in the Hard X-ray Pulses
from the Magnetar 1E 1547.0$-$5408
}

\author{
Kazuo  \textsc{Makishima},\altaffilmark{1,2,3}
Teruaki \textsc{Enoto},\altaffilmark{4,5,6}
Hiroaki \textsc{Murakami},\altaffilmark{1}
Yoshihiro \textsc{Furuta},\altaffilmark{1}
Toshio \textsc{Nakano},\altaffilmark{1,2}
Makoto \textsc{Sasano},\altaffilmark{1}
and
Kazuhiro \textsc{Nakazawa}\altaffilmark{1,2}
}

\email{maxima@phys.s.u-tokyo.ac.jp}

\altaffiltext{1}{
Department of Physics, The University of Tokyo,
7-3-1 Hongo, 
Bunkyo-ku, Tokyo 113-0033, Japan
}
\altaffiltext{2}{
Research Center for the Early Universe,
The University of Tokyo, \\
7-3-1 Hongo, Bunkyo-ku, Tokyo 113-0033, Japan
}
\altaffiltext{3}{
MAXI Team, 
The Institute of Physics and Chemical Research (RIKEN),\\
2-1 Hirosawa, Wako, Saitama 351-0198, Japan
}
\altaffiltext{4}{
High Energy Astrophysics Laboratory,
RIKEN Nishina Center,\\
2-1 Hirosawa, Wako, Saitama 351-0198, Japan
}
\altaffiltext{5}{
 The Hakubi Center for Advanced Research, Kyoto University, 
 Kyoto 606-8302, Japan}
\altaffiltext{6}{
Department of Astronomy, Kyoto University, Kitashirakawa-Oiwake-cho, 
Sakyo-ku, Kyoto 606-8502, Japan
}

%
\maketitle

\begin{abstract}
The Suzaku data of the highly variable magnetar 1E 1547.0$-$5408,
obtained during the 2009 January activity, were reanalyzed.
The 2.07 s pulsation,
detected in the 15--40 keV HXD data,
was found to exhibit phase modulation,
which can be modeled by a sinusoid 
with a period of $36.0^{+4.5}_{-2.5}$  ks and an amplitude of $0.52 \pm 0.14$ s.
While the effect is also seen in the 10--14 keV XIS data,
the modulation amplitude decreased towards lower energies,
becoming consistent with 0 below 4 keV.
After the case of 4U 0142+61,
this makes the 2nd example of this kind of behavior 
detected from magnetars.
The effect can be interpreted as a manifestation of 
torque-free precession of this magnetar,
which is suggested to be prolately deformed
under the presence of strong toroidal field of $\sim 10^{16}$ G.
\end{abstract}

\section{Introduction}
\label{sec:intro}
Magnetars (e.g., \cite{Magnetar,Mereghetti08}) are considered 
to be magnetically-powered isolated neutron stars (NSs) 
having  ultra-strong dipole magnetic fields, $B_{\rm d} = 10^{14-15}$ G,
as derived from their pulse periods $P$ and period derivatives $\dot{P}$.
These values of $B_{\rm d}$ exceed so-called critical field, $4.4 \times 10^{13}$ G,
above which the Landau level separations of electrons exceed their rest-mass energy.
Magnetars emit characteristic two-component X-ray spectra,
consisting of the long-observed soft component
and the mysterious hard component \citep{Kuiper06,denHartog06,Enoto10b}.

Magnetars are thought to harbor even higher 
{\it toroidal} (or internal) magnetic fields $B_{\rm t}$
(e.g., \cite{Magnetar,Harding+Lai06,toroidal09b,toroidal09a}).
These fields may be generated and amplified
in the progenitor's core,
during its final gravitational collapse 
involving strong differential rotation (e.g., \cite{Takiwaki09}).
Some models explain the production of energetic particles 
in magnetars invoking toroidal magnetic fields 
and the associated twisted magnetic configuration 
(e.g., \cite{Twist07}).

Observationally, the presence of intense $B_{\rm t}$ is supported
by the recently discovered low-$B_{\rm d}$ magnetars, 
SGR 0418+5729 \citep{Rea13},
SGR1822$-$1606 \citep{Rea12},
and 3XMM J185246.6+003317 \citep{Rea14};
their burst activity suggests that their internal fields
are much stronger than  their externally observable fields of
$B_{\rm d}=6 \times 10^{12}$ G,  $2.7 \times 10^{13}$ G, 
and $<1.4 \times 10^{13}$ G,
respectively.
The possible proton cyclotron feature in SGR 0418+5729,
which suggests a local surface field of $>2 \times 10^{14}$ G 
\citep{Tiengo13}, provides another support,
because a part of such intense internal fields would emerge from 
the NS surface to form local multipoles.
Thus, the strong $B_{\rm t}$ may be regarded as
one of the most important properties of magnetars.
However, it remained  difficult to more directly estimate $B_{\rm t}$,
because toroidal fields are intrinsically  confined inside the NS
and invisible from outside.

After pioneering works by, e.g., \citet{Treves} and \citet{Heyl02},
the way around this difficulty was explored by 
Makisihma et al. (2014; hereafter MEA14)  in the following way.
When an NS has  $B_t \sim 10^{16}$ G,
is it expected to be deformed into a prolate shape \citep{Ioka01, Cutler02,Ioka+Sasaki04},
to a degree of
\begin{equation}
\epsilon \equiv (I_1 - I_3)/I_3  \sim 1\times 10^{-4} (B_{\rm t}/10^{16}{\rm G})^2
\label{eq:epsilon}
\end{equation}
where $I_3$ is the moment of inertia  
around  the NS's symmetry axis $\hat{x}_3$,
and $I_1$  that around axes orthogonal to $\hat{x}_3$.
Under the absence of external torque,
such an NS has a constant rotation period 
$P_{\rm rot} \equiv 2\pi I_3/L$ around $\hat{x}_3$,
and a constant precession period $P_{\rm prec} \equiv 2\pi I_1/L$ 
around the angular momentum vector $\vec{L}$ 
(with $L \equiv |\vec{L}|$) which is fixed to the inertial frame.
These two periods, which were degenerate when spherical,
thus become different just by $\epsilon$.
If  $\hat{x}_3$ is tilted from $\vec{L}$ by 
a non-zero {\it wobbling angle}  $\alpha$,
the NS is said to exhibit  {\it free precession},
which is the most basic behavior of a torque-free 
rigid body with an axial symmetry \citep{Landau+Lifshitz, Butikov06}.
This concept should be distinguished from the more familiar {\it forced precession}
driven by external torque,
as seen in spinning tops.

Suppose that a precessing NS has a radiation pattern 
that  breaks axial symmetry around $\hat{x}_3$.
Then, as argued by  MEA14,
the observed pulses will be phase-modulated at 
the beat period between  $P_{\rm rot}$ and $P_{\rm prec}$,
which is  called {\it slip period} and  is given as
\begin{equation}
T \equiv P_{\rm prec}/\epsilon = (1/P_{\rm rot} - 1/P_{\rm prec})^{-1}~.
\label{eq:slip}
\end{equation}
Using the Suzaku X-ray observatory \citep{Suzaku},
MEA14 in fact found that the $P=8.687$ s pulsation 
of the spectral hard component of the magnetar 4U 0142+61 is 
subject to  phase modulation with a  period of $T=55 \pm 4$ ks.
When this effect is interpreted as a manifestation of 
free precession and $T$ is identified with the associated slip period,
equation (\ref{eq:slip}) yields $\epsilon= 1.6 \times 10^{-4}$,
and hence equation (\ref{eq:epsilon}) implies $B_{\rm t} \sim 10^{16}$ G.
This provides one of the first direct estimations of $B_{\rm t}$ of a magnetar.

As seen above,
the essential assumptions made by MEA14
to interpret the observed phase modulation of 4U~0142+61 are;
$\epsilon \neq 0$, $\alpha \neq 0$,
and the broken axial symmetry of the hard X-ray emission pattern around $\hat{x}_3$.
If these assumptions generally hold in objects of this class,
we should observe the same effects from other magnetars as well.
If detected from a fair number of them,
the effect will provide such important information as;
how $B_{\rm t}$ is distributed among them,
whether $B_{\rm d}$ and $B_{\rm t}$ are correlated,
and whether $B_{\rm t}$ actually decreases 
(as postulated in the magnetar scenario itself; \cite{Magnetar})
towards objects with larger characteristic ages.
The results will also provide valuable clues to the origin
of the hard X-ray component \citep{Kuiper06,denHartog06},
to the nature of the intriguing spectral evolution 
of magnetars discovered by \citet{Enoto10b},
and to the ultimate origin of these mysterious objects
(e.g., \cite{Vink06,Nakano15}).
Thus,  examining other magnetars for the hard X-ray phase modulation
is extremely important and urgent.

In the present paper,
we have chosen the highly variable magnetar 1E 1547.0$-$5408
as our 2nd target.
It is the fastest-rotating magnetar with $P=2.07$ s,
possibly associated with the supernova remnant
G327.24$-$0.13 \citep{Gelfand+Gaensler}.
The observed value of $\dot{P} = 2.4 \times 10^{-11}$s s$^{-1}$ \citep{Kuiper12},
together with $P$, yields $B_{\rm d} = 2.2\times 10^{14}$ G
and a very young characteristic age of $\tau \equiv P/2\dot{P} = 1.4$ kyr.
On 2009 January 22,  a strong bursting activity 
was detected from  1E 1547.0$-$5408 with Swift (Gronwall et al. 2009). 
This activity was continuously covered by RXTE, INTEGRAL, 
and Swift observations \citep{Kuiper12}.
These monitoring studies revealed  
an increase in the total flux  at the burst onset
by more than an order of magnitude,
followed by a rapid decline,
and complex changes in the spin-down characteristics
along the activity evolution.

As already reported by \citet{Enoto10a} (hereafter Paper I) and \citet{Enoto12},
this object was observed by Suzaku during its 2009 January outburst,
and yielded by far the strongest hard X-ray ($>10$ keV) signals 
among some 10 magnetars so far observed with Suzaku.
Although this source was observed with Suzaku again on 2010 August,
these follow-up data are not utilized here,
since the source had faded by that time in hard X-rays  to a level 
comparable to the HXD detection limit \citep{Iwahashi13}.

\section{Observation }

As described in  Paper I and \citet{Enoto12},
the present Suzaku observation of 1E 1547.0$-$5408 was
performed  from 2009 January 28 21:34 (UT) to January 29 21:32 (UT),
for a gross duration of  86 ks,
with the target placed at the HXD nominal position.
This was a Target-of-Opportunity observation
in response to the activity onset,
and this epoch falls on ``segment 3" 
(from 2009 January 22 to February 1) of \citet{Kuiper12}.
Among the three operating cameras of 
the X-ray Imaging Spectrometer (XIS: \cite{XIS}), 
XIS0 was operated in the timing mode (P-sum mode), 
with a 7.8 ms time resolution.
The Hard X-ray Detector (HXD: \cite{HXD}) was
operated in the standard mode, 
yielding a  time resolution of 61$~\mu$s. 
Further details of the observation are found in Paper I.

Below, we mainly utilize the data from XIS0 and HXD-PIN
 processed as described in Paper I,
 in which each detected event is tagged with the energy and the arrival time.
Those from XIS1 and XIS3
had too low a  time resolution (2.0 s) to resolve the pulsation,
and signal statistics were rather low in HXD-GSO.
The net exposures achieved with XIS0 was 42.6 ks,
while that with the HXD  was 33.5 ks.
After subtracting the background and applying dead-time corrections,
the object was detected with HXD-PIN
at a 15--70 keV count rate of $0.174 \pm 0.004$ c s$^{-1}$ (Paper I),
and a 1--10 keV rate of  $5.5 \pm 0.1$ c s$^{-1}$ with XIS0.

As described in Paper I, 
13 short-burst candidates were removed,
and arrival times of the XIS0 and HXD-PIN events were photon-by-photon
converted  to those to be measured at the solar-system barycenter.
The conversion utilized  the source position of $(\alpha^{2000}, \delta^{2000})
= (15^{\rm h}50^{\rm m}54^{\rm s}.11,  54^\circ18'23.''7)$,
as well as  the spacecraft orbital information.

\section{Data analysis and results}

\subsection{Reconfirmation of Paper I}

Since the  present paper  re-analyzes  the same data set  
of 1E 1547.0$-$5408 as used in Paper I,
let us begin with reproducing the timing results of that publication.
While Paper I utilized the conventional chi-square method
to search for the pulsations,
here we instead employ the $Z_n^2$ technique \citep{Zn2_83,Zn2_94}
which is free from the ambiguity of phase-bin number  per pulse cycle
and is generally less affected by the Poisson errors.
For each trial pulse period $P$,
this technique effectively calculates the folded pulse profile
using a sufficiently large bin number (in practice not specified).
Then, the Fourier power of the profile is summed up to 
a specified harmonic number  $n~(n=1, 2, ...$),
and normalized in a certain manner to reflect the data statistics.
This process yields the statistical quantity called $Z_n^2$,
which evaluates the significance of periodicity at $P$.
If no intrinsic periodicity is present at $P$ except  Poissonian fluctuations,
the $Z_n^2$ value calculated according to the normalization of \citet{Zn2_94},
which we adopt here,
should obey a chi-square distribution with $2n$ degrees of freedom.
The higher noise tolerance of  this  technique is realized by
discarding the power in all harmonics higher than is specified.

\begin{figure}[htb]
  \begin{center}
   \FigureFile(80mm,){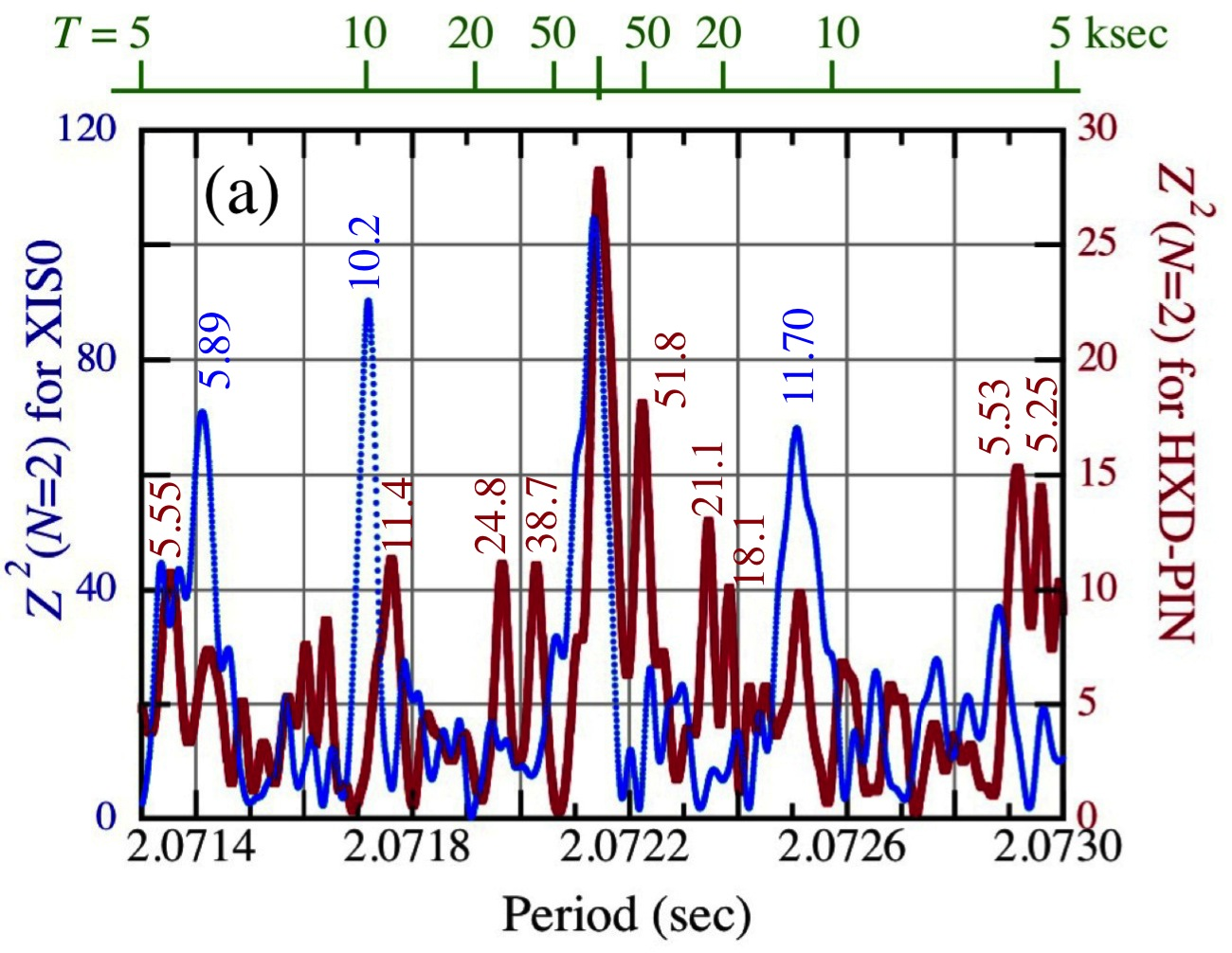}
    \FigureFile(80mm,){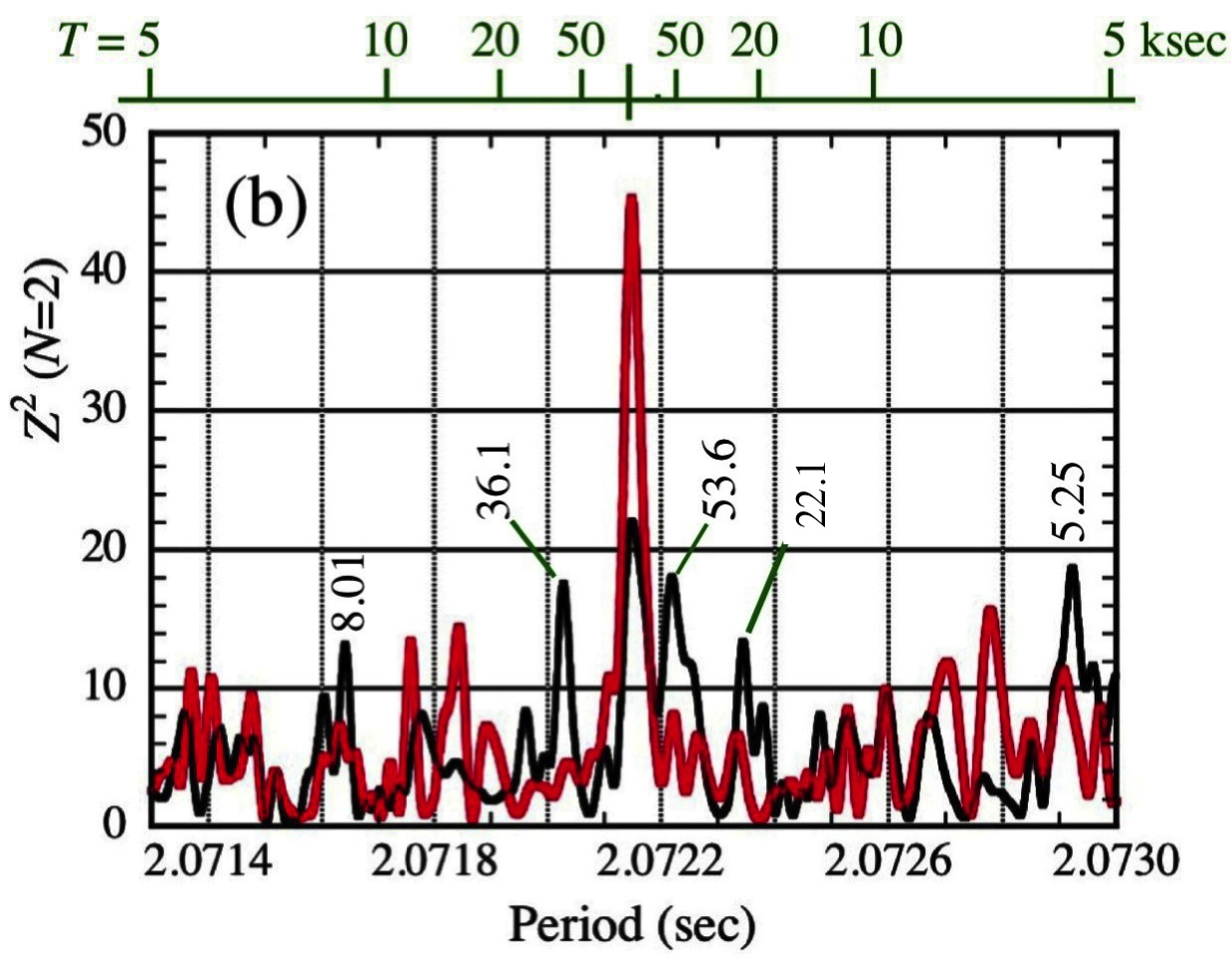}
  \end{center}
   \caption{(a) Periodogams of 1E 1547.0$-$5408 
  calculated using the $Z_n^2$ method with $n=2$,
from the 2--10 keV XIS0 data (blue) and those of HXD-PIN in 12--70 keV (brown).
(b) The same as panel (a), but for the 15--40 keV HXD-PIN data,
before (black) and after (red) applying the demodulation analysis
employing the parameters of equation (\ref{eq:best_para_HXD}).
In both panels, the green abscissa at the top indicates the period $T$
which can explain a side lobe at that period
in terms of its beat with the pulsation.
The implied values of $T$ (in units of ks)  of major side lobes
are also indicated in the figure.
}
 \label{fig:periodogram0}
  \end{figure}

Figure~\ref{fig:periodogram0}a shows periodograms calculated 
in this way using the  2--10 keV XIS0 data and the 12--70 keV HXD-PIN data,
both including backgrounds.
The harmonic number was chosen to be $n=2$,
considering the double-peaked pulse profile of this source (e.g., Paper I).
These energy bands, as well as the trial period range (2.0713--2.0730 s),
are the same as those  in Paper I,
while the employed period search step, $1 \times 10^{-6}$ s,
is 4 times finer.
The results reconfirm essential features of figure 3 of Paper I, 
including the best-estimate period of
\begin{equation}
P_{\rm 0s} = 2,072.135 \pm 0.005~{\rm ms}
\label{eq:P0_XIS}
\end{equation}
referring to the XIS result,
and the side lobe patterns.
The HXD periodogram peak,
\begin{equation}
P_{\rm 0h} = 2,072.147 \pm 0.005~{\rm ms}~,
\label{eq:P0_HXD}
\end{equation}
is consistent with equation (\ref{eq:P0_XIS}) within errors,
although it is slightly displaced to longer periods.
This behavior is also seen in figure 3 of Paper I.
The errors in $P_{\rm 0s}$ and $P_{\rm 0h}$ 
refer to the range
where pulse-phase differences  amount to  $\pm 10\%$ of one pulse cycle
when accumulated over the gross duration of 86 ks.
These errors approximately agree with widths of the peaks in the periodograms.

At $P_{\rm 0h}$, 
the 12--70 keV HXD periodogram in figure~\ref{fig:periodogram0}a
reaches the peak height of $X =28.21$,
where $Z_2^2$ is abbreviated to $X$.
In terms of chi-square distribution with $2n=4$ degrees of freedom,
the chance probability of finding $X$ values higher than this
is found as  ${\cal P}_{2}^{\rm h} = 1.1 \times 10^{-4}$;
here, the subscript specifies the harmonic number $n$,
and the superscript ``h"means the HXD.
This agrees with the probablity of $\sim 1 \times 10^{-4}$
obtained in Paper  I  in 10--70 keV with the chi-square technique.

In figure~\ref{fig:periodogram0}a, 
the XIS and HXD periodograms both show a number of side lobes.
Such a structure at a period $P'$ often arises
due to  beat between the pulsation at $P_0$ 
(representing $P_{\rm 0s}$ or $P_{\rm 0h}$)
and a certain longer period $T$.
Based on this assumption,
we calculated the required period as $T=|(1/P_0 -1/P')^{-1}|$,
and show it in green at the top of figure~\ref{fig:periodogram0}.
In addition,  the required values of $T$ (in units of ks) of 
major XIS0 and HXD side lobes are given explicitly in the figure.
Thus, many of the side lobes (particularly of XIS0) in figure~\ref{fig:periodogram0}a
can be explained as a beat of  $P_0$ with  $T=5- 6$ ks or $T=10-12$ ks,
which agree with the Suzaku's orbital period $P_{\rm orb} = 5.6$ ks
or twice this value, respectively.
However, the HXD-PIN periodogram shows
several unexplained peaks;
at $T=24.8$ ks and 38.7 ks on the left side of  $P_0$,
as well as at $T=51.8$ ks,  21.1 ks, and 18.1 ks on the right side.
These structures, also seen in figure 3 of Paper I, 
suggest that the hard X-ray pulses could be affected
by some long periodicity in the range of $T=18-52$ ks (5--14 hr).

\subsection{Detailed analysis of the HXD data}
\label{subsec:HXD}

Below, we analyze the HXD-PIN data in further details, 
selecting an energy range of 15--40 keV.
The lower energy bound  is thus raised
from 12 keV used in figure~\ref{fig:periodogram0}a to 15 keV,
to avoid increased soft thermal noise.
Likewise, the upper energy bound is lowered from 70 keV to 40 keV,
to reduce the contribution from the non X-ray background.

\subsubsection{Evolution of hard X-ray pulse profiles}
\label{subsec:Pr_evol}
The  $Z_2^2$ periodogram calculated from the 15--40 keV HXD-PIN data
is shown in figure~\ref{fig:periodogram0}b in black.
Except some small differences,
it  is similar to the 12--70 keV results shown in brown in panel (a).
The peak value, also obtained at the period of equation~(\ref{eq:P0_HXD}), 
becomes
\begin{equation}
X_0 \equiv (X)_{A=0} = 22.06  ~,
\label{eq:Z2_0}
\end{equation}
which is somewhat smaller than that in 12--70 keV.

In the 15--40 keV periodogram,
the period range of $T=18-52$ ks,
where the unexplained side lobes were noticed in figure~\ref{fig:periodogram0}a,
is approximately preserved (although detailed of the side lobes changed).
Since this period range does not contain particular artificial periodicities
of the observation such as data gaps or background variations,
some properties (e.g., amplitude, phase, or pulse shapes)
of the 2.07 s hard X-ray pulsation could be subject to 
slow variations with periods in this range.
In fact,  \citet{Kuiper12} found that
the pulse shapes were changeable during their ``segment 3"
which covers the present Suzaku observation,
that  a new intermediate peak in the $>4$ keV pulse profile emerged,
and that  the pulsed hard X-ray flux {\it increased} meantime
although the total 4--27 keV flux was rapidly declining.

To examine the evolution of the hard X-ray pulse properties,
we divided the whole observation into consecutive 
time segments of approximately equal lengths, 
denoted as $j=0, 1, ..., J-1$
($j=0$ at the beginning and $j=J$ at the end),
and sorted the entire 15--40 keV HXD-PIN events
(including backgrounds)  
into a two-dimensional array $C(j,k)$.
Here,  the column $k=0, 1.., K-1$ represents the pulse phase
(i.e., the photon arrival times modulo $P_{\rm 0h}$).
Thus, a set $\{C(j,k); k=0,.., K-1\}$ gives 
the pulse profile in the $j$-th time segment.
We employed $J=6$ and $K=12$,
and show in figure~\ref{fig:6pulse_profiles}a 
the obtained  time-divided six pulse profiles
as a function of $k$.
There, running averages were applied to 
3 consecutive data points $(k-1, k, k+1)$.
Thus, the pulse amplitude changes with time,
and the pulse shape is also variable;
it is relatively flat-topped with a single deep minimum,
with an occasional shallower secondary minimum at about half a cycle offset.
Such a short-term variation in the pulse shape 
was observed from 4U 0142+61 \citep{Enoto11},
as well as from 1E~1547$-$5408 itself \citep{Kuiper12}.

More importantly,  
thanks to the high sensitivity of the Suzaku HXD 
on relatively short time scales \citep{HXD2},
the pulse {\it phase} is  seen to vary 
clearly in figure~\ref{fig:6pulse_profiles}a,
by up to about half a pulse cycle.
On the other hand,
we do not notice in figure~\ref{fig:6pulse_profiles}a 
particular {\it secular} changes in the pulse profile, pulsed flux, or pulse phase.
This does not  contradict to the discovery by \citet{Kuiper12}
of the emerging new component in the hard X-ray pulse profile,
because the growth of that component is
estimated to be at most $\sim 10\%$ across the present Suzaku pointing,
as judged from the history of the hard X-ray pulsed flux 
(figure 9 of \cite{Kuiper12}).

\subsubsection{Periodic phase modulation}
\label{subsec:periodic_modulation}
We next need  to know
whether the pulse-phase fluctuations seen in  
figure~\ref{fig:6pulse_profiles}a are {\it periodic} or not.
This requires a finer time division by increasing $J$,
but  individual pulse profiles would then lose statistics.
Therefore, we chose another way;
the data accumulation was repeated 
with $J=20$ and $K=5$,
but the results are presented in figure~\ref{fig:5ltcvs}a
in the form of five {\it light curves},
$\{{C(j,k); j=0,.., J=19}\}$ using $j$ as abscissa.
For presentation,  
3-point running averages were again applied
over  $(j-1, j, j+1)$.
Thus, the background-inclusive count rates 
in the five pulse phases mostly varied rather periodically, 
by  2 to 3 cycles  across the 86 ks gross pointing,
although the amplitude is comparable to the  statistical errors.

We calculated discrete Fourier power spectra 
of the five light curves in figure~\ref{fig:5ltcvs}a,
before taking the running averages,
and summed them to obtain a single power spectrum
which is shown by a red solid line in figure~\ref{fig:5ltcvs}c.
It indeed reconfirms the suggested  power enhancement,
at  wave numbers of $q=2$ and $q=3$,
with the corresponding period range of
\begin{equation}
 86/3= 29~{\rm ks~~~to~~} 86/2 = 43~{\rm ks}~.
 \label{eq:T-range}
 \end{equation}
The weighted mean of $q=2$ and $q=3$ 
over the power spectrum in figure~\ref{fig:5ltcvs}c becomes
\begin{equation}
q = 2.45,~~~{\rm or} ~~T=35.1 ~ {\rm ks}~,
\label{eq:best_wavenum}
\end{equation}
which is fully consistent with the implication of the side lobes
in figure~\ref{fig:periodogram0}.

As a ``control" study,
figure~\ref{fig:5ltcvs}b presents another set of light curves,
calculated in the same way 
but purposely changing the folding period
from $P_0$ to a dummy value of 2.0 s.
Its behavior is clearly more random,
without  particularly favored periods.
Indeed, these 5 control light curves yielded
the blue power spectrum in figure~\ref{fig:5ltcvs}c,
which is  consistent with white noise.

To evaluate the significance of the $q=2$ and $q=3$ 
enhancements in the red power spectrum of figure~\ref{fig:5ltcvs}c,
we repeated the same control study 20 times,
by changing the dummy period around 2.0 s.
The derived ensemble-averaged power spectrum,
shown in figure~\ref{fig:5ltcvs}c by a dashed black line,
is approximately white
(with hints of red noise at $q=1$ and background variation at $q=4$
corresponding to $5 P_{\rm orb}$).
Around this mean,
individual power spectra exhibited a $1\sigma$ scatter by $\sim \pm 0.5$.
Therefore, the $q=2$ and $q=3$ excess (above 1.0)  
in the red power spectrum is estimated to be 2.4 and 1.6 sigma effects,
and their sum implies a $\sim 3$ sigma excess.

Since the five light curves in figure~\ref{fig:5ltcvs}a do not vary in phase,
the suggested long periodicity is unlikely to result from pulse-amplitude variations,
but  considered to reflect the pulse-phase jittering
found in figure~\ref{fig:6pulse_profiles}a.
Thus, the hard X-ray pulsation is inferred to  
exhibit {\it periodic phase modulation}.
To reinforce this inference,
we accumulated the pulse profiles
in 6 separate phases of an assumed periodicity of $T=36$ ks,
of which the choice  is justified later.
The result is presented in figure~\ref{fig:6pulse_profiles}b;
in this case, the row $j$ of the array $C(j,k)$ denotes
the pulse phase, i.e., the arrival times modulo $T$, 
rather than the arrival times themselves.
It visualize a periodic and approximately sinusoidal pulse-phase modulation,
with an amplitude of
$\Delta k \sim \pm 3.5$ bins
($\pm 0.3$  pulse cycle or $\pm 0.6$ s)
which is consistent with the inference from figure~\ref{fig:6pulse_profiles}a.

\subsubsection{Demodulation analysis}
\label{subsec:demod_HXD}

The above semi-quantitative analysis justifies us
to apply the  ``demodulation analysis",
developed by MEA14, to  the HXD data,
to more rigorously quantify the pulse-phase modulation.
Specifically, we consider that the hard X-ray pulsation of 
1E 1547.0$-$5408 is somehow phase modulated,
like in 4U~0142+61, 
at a period $T$ in the range of equation (\ref{eq:T-range}).
We hence hence assume
that the peak timing $t$ of each  2.07 s pulse is displaced sinusoidally by
\begin{equation}
\Delta t = A \sin (2\pi t/T -\phi)~,
\label{eq:modulation}
\end{equation}
where $A$  and $\phi$ are  the amplitude and initial phase 
of the assumed modulation, respectively.

As formulated in MEA14, such effects can be removed
by shifting the arrival times of individual HXD photons  back by $-\Delta t$.
This procedure was previously employed, e.g.,  by \citet{Koyama89} 
to search for orbital motions in the anomalous X-ray pulsar 1E 2259+586,
and by \citet{Terada08} to remove the orbital effects 
in the fastest-spinning accreting white dwarf binary, AE Aqr,
and successfully detect its hard X-ray pulsations.
We hence applied these time displacements to the individual HXD photons,
and re-calculated  the $Z_2^2$ periodograms
to see whether  the pulse significance changes.
The pulse period was scanned over a range of
$P=2.072130-2.072152$ (with a step $1~\mu$sec),
i.e., the joint error range of 
equation (\ref{eq:P0_XIS}) and equation (\ref{eq:P0_HXD}).
The  triplet $(T, A, \phi$) describing the phase modulation was 
scanned over $A=0-0.80$ sec (with a $0.05$ sec step), 
$\phi=0-360^\circ$ ($3^\circ-10^\circ$ step),
and $T=10-80$ ks (with a step of 0.2 to 0.5 ks).
The harmonic number $n=2$ is retained.

\begin{figure}[htb]
  \begin{center}
   \FigureFile(110mm,){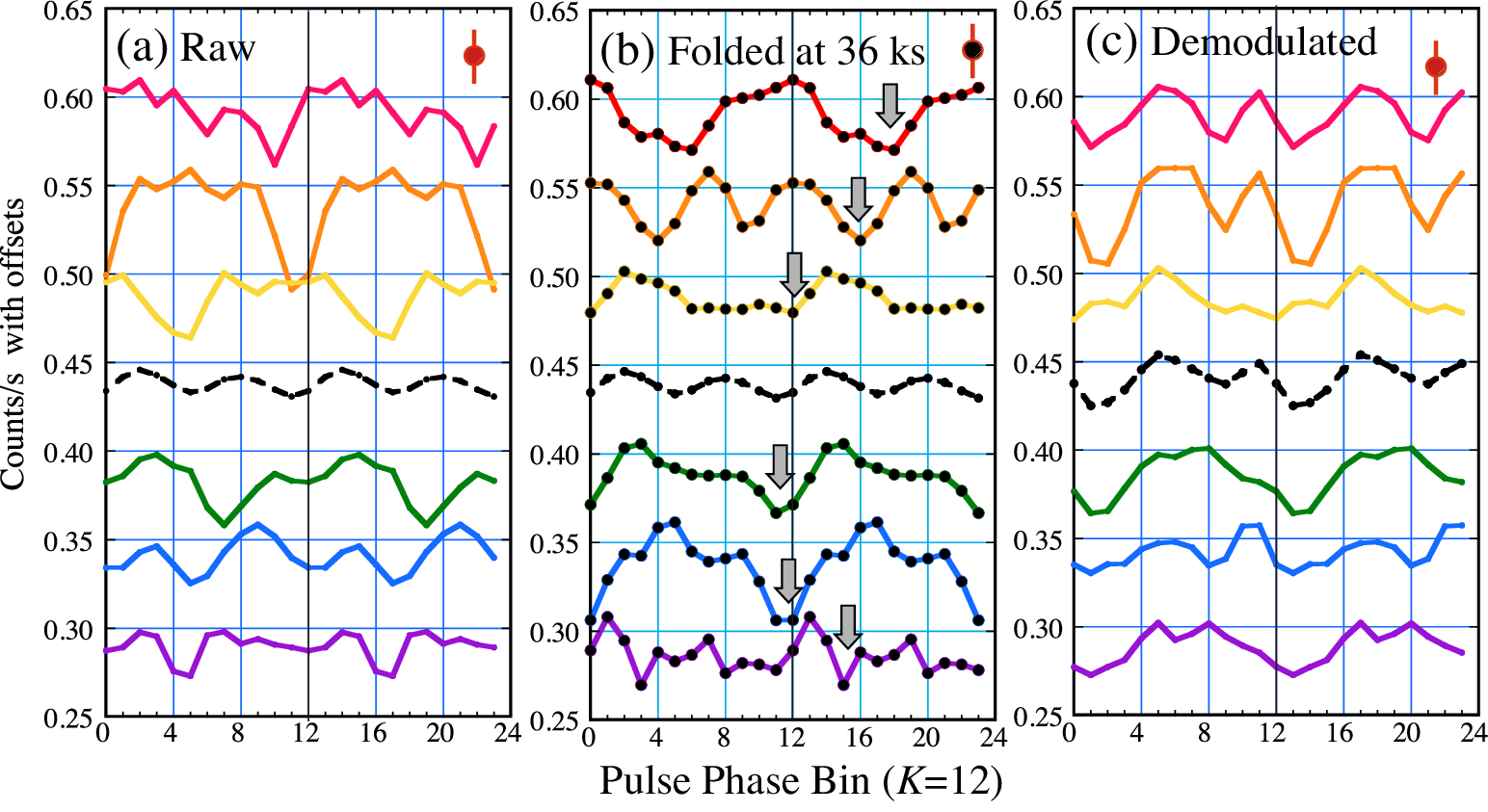} 
 \FigureFile(50mm,){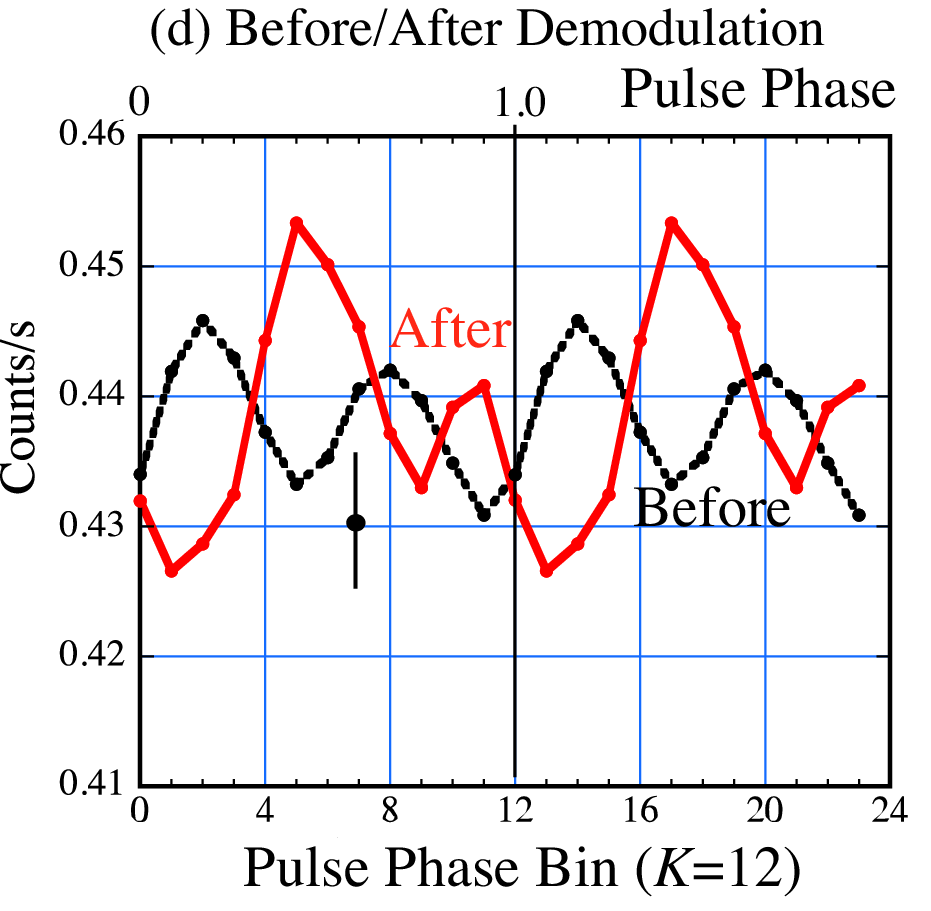}
   \end{center}
   \caption{
   (a) The 15--40 keV background-inclusive pulse profiles
   folded at equation (\ref{eq:P0_HXD}),
  obtained in 6 consecutive time segments of $\sim 14$ ks each
  (from red to purple).
  They are shown for two cycles, with vertical offsets,
   after applying running averages over 3 adjacent bins.
   A typical $\pm 1\sigma$ statistical error is given at the top right corner.
  The dashed black profile at the middle represents the average.
  (b) The 15--40 keV folded pulse profiles
accumulated over six different phases
   of the  $T=36$ ks modulation period, shown with offsets.
   Meaning of the colors is different from that in panel (a).
   A possible phase assignment is indicated by arrows.
  (c) The same as panel (a), but prior to the folding,
  the arrival time of each event was corrected with equation (\ref{eq:modulation})
  using the parameters of equation (\ref{eq:best_para_HXD}).
  (d) A comparison between the 15--40 keV pulse profiles
  before (black) and after (red) the demodulation, both with the running averages.
  The former is the same as the black profiles in panels (a) and (b),
  while the latter is identical to the black one in panel (c).
  The background level is at about 0.27 c s$^{-1}$.
}
 \label{fig:6pulse_profiles}
  \end{figure}

\begin{figure}[htb]
  \begin{center}
    \FigureFile(58mm,58mm){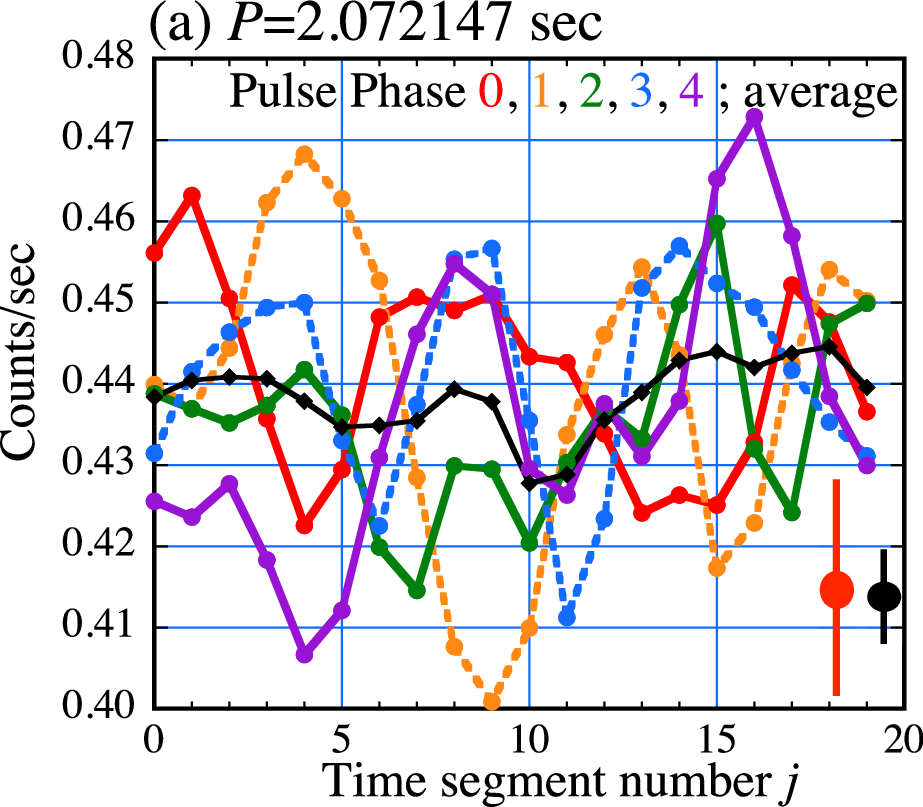}
   \FigureFile(58mm,){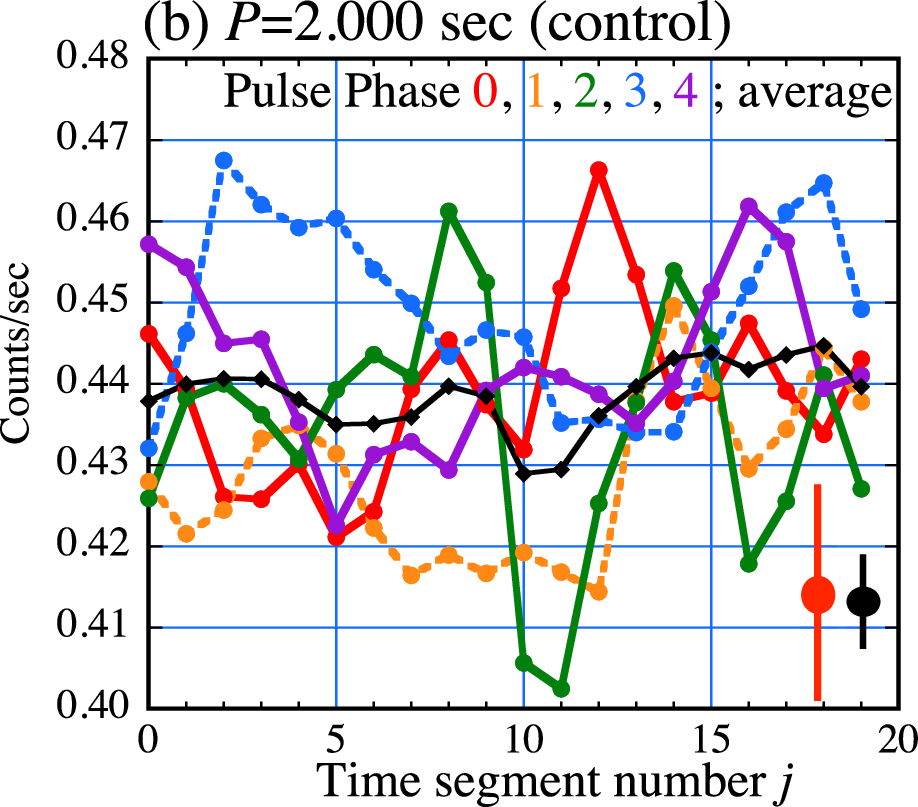}
   \FigureFile(42mm,){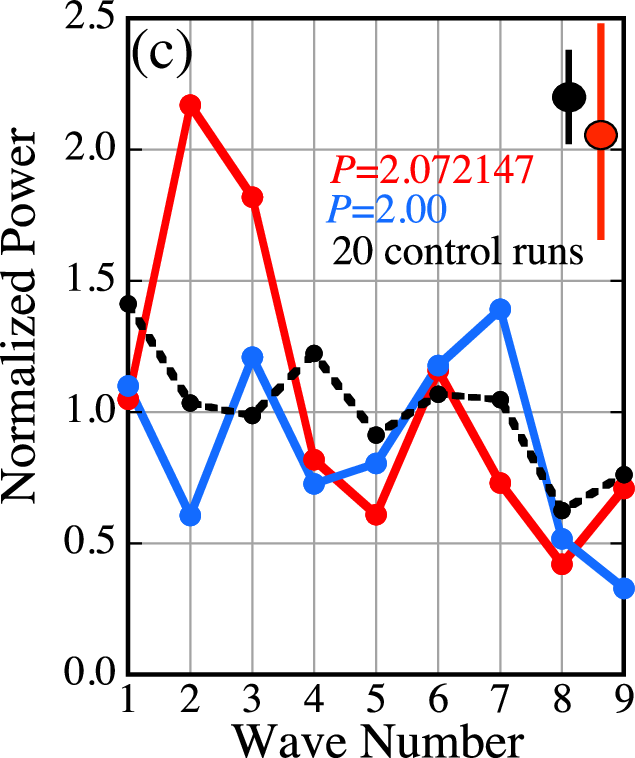}
  \end{center}
   \caption{
  (a) Background-inclusive 15--40 keV  light curves of 1E 1547.0$-$5408
  each covering the entire observation with 20 segments ($j=0, 1, .., 19$),
  derived in 5 different phases  (red, orange, green, blue, and purple)
  of the pulse period of  equation (\ref{eq:P0_HXD}).
  See text for details.
  They are shown after applying a running average over the adjacent 3 bins,
  and their average is given in  solid black line.
  Typical $\pm 1\sigma$ error bars are given at the bottom right corner.
  (b) The same as panel (a), but sorted into 5 phases of a dummy period of 2.0 s.
  (c) Power spectra of the five light curves in panels (a) and (b),
 in red and blue lines, respectively.
  Both are normalized to the power expected from the
  Poisson fluctuations.
 The black dashed line represents an ensemble average 
  over 20 control studies, each similar to panel (b).
}
 \label{fig:5ltcvs}
  \end{figure}

\begin{figure}[bht]
  \begin{center}
   \FigureFile(140mm,){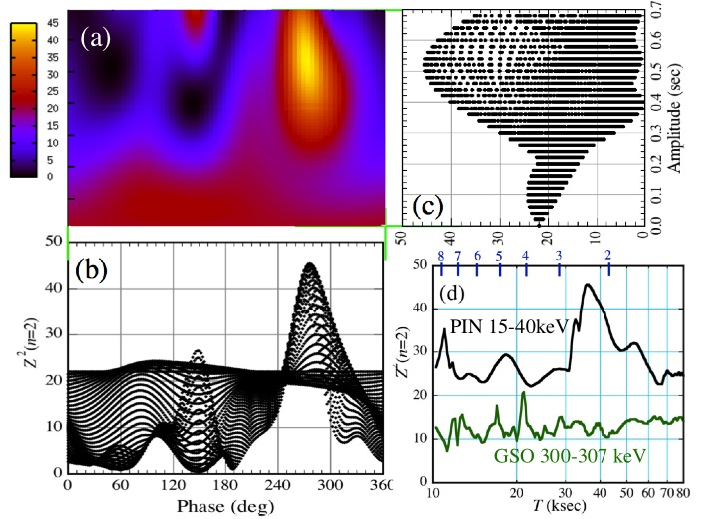}
  \end{center}
 \caption{Results of the demodulation analysis on the 15--40 keV HXD-PIN data,
  assuming a periodic phase modulation described by equation (\ref{eq:modulation}).
(a)  A two-dimensional color map
of $X \equiv Z_2^2$ (maximum as $P$ is varied) 
on the $(A, \phi)$ plane, assuming $T=36.0$ ks.
(b) Projection of panel (a) onto the $\phi$ axis.
(c) Projection onto the $A$ axis.
(d) $T$-dependence of the $Z_2^2$ peak value, $X_{\rm pk}(T)$,
presented in black.
Green shows a reference result using the 300--307 keV HXD-GSO data.
The corresponding Fourier wave number,
for the total time span of 86 ks, is shown in blue at the top.
  }
 \label{fig:demodulation_HXD}
  \end{figure}

The results of this {\it demodulation} analysis are 
presented in figure~\ref{fig:demodulation_HXD},
just in the same format as figure 2 of MEA14.
There, panel (a) shows a two-dimensional color map
of the  value of $ X \equiv Z_2^2$ (the maximum as $P$ is varied) 
obtained on each grid point $(A, \phi)$, 
for a particular case  of $T=36.0$ ks.
Its projections on the $\phi$ and $A$ axes are
presented in panels (b) and (c), respectively,
where multiple traces indicate cross sections of panel (a) at various positions.
While the value of $X_0$ in equation (\ref{eq:Z2_0}) is reproduced at $A=0$ in panel (c),
a clear peak is seen at $A\sim 0.52$ s and $\phi \sim 270^\circ$.
This value of $A$ agrees well with what can be read from figure~\ref{fig:6pulse_profiles}b.
Furthermore, the value of $\phi$, together with equation (\ref{eq:modulation}),
predicts $\Delta t = +A$ (the maximum delay) at about
the beginning $(t\sim 0)$ of the $T=36$ ks phase, 
again in agreement with figure~\ref{fig:6pulse_profiles}b.
Although figure~\ref{fig:6pulse_profiles}a apparently
suggests a rather long periodicity (e.g., $T>100$ ks),
this is probably a beat between the $T=36$ ks modulation
and the Nyquist period therein ($1 \times 86/6=29$ ks),
which should appear at 140 ks.

In panels (a)-(c) of figure~\ref{fig:demodulation_HXD},
the peak significance reaches $X_{\rm pk}= 45.60$;
here,  $X_{\rm pk}(T)$ denotes the peak value of $X$
found on a  $(A, \phi)$ plane for a given value of $T$.
Panel (d) of figure~\ref{fig:demodulation_HXD} shows 
how this $X_{\rm pk}(T)$ depends on the assumed modulation period $T$.
Over the entire $(P, T, A, \phi)$ space surveyed,
the highest value of $X$,
to be called the grand-maximum
and denoted $X_{\rm max}$, is thus found as 
\begin{equation}
X_{\rm max} \equiv X_{\rm pk}(36.0{\rm ks}) = 45.60  ~,
\label{eq:Z2_max}
\end{equation}
for a set of modulation parameters given as
\begin{equation}
A=0.52 \pm 0.14~{\rm s}, ~~ \phi= 276^\circ \pm 15^\circ, ~~~ T=36.0^{+4.5}_{-2.5}~{\rm ks}~.
\label{eq:best_para_HXD}
\end{equation}
Here, the errors are represented in the same way as in MEA14,
i.e., by the standard deviations of  Gaussians
fitted to the distributions (above uniform backgrounds)
in figure~\ref{fig:demodulation_HXD}.
This value of $T$,
already employed to produce figure~\ref{fig:6pulse_profiles}b,
 agrees well with equation (\ref{eq:best_wavenum}).

As shown in figure \ref{fig:periodogram0}b in red,
the demodulation with equation (\ref{eq:best_para_HXD}) 
has changed the HXD-PIN periodogram significantly.
The peak  at $P_{\rm 0h}$ has become much higher,
just from $X_0$  to  $X_{\rm max}$.
Furthermore, the  $P_{\rm orb}$-unrelated side lobes have diminished;
powers therein are thought to have been restored to the main peak.

In figure~\ref{fig:6pulse_profiles}c,
we again calculated the 15--40 keV folded pulse profiles
in the same 6 time segments as in figure~\ref{fig:6pulse_profiles}a,
but after applying the time corrections by equation (\ref{eq:modulation})
to individual photons
using the parameters in equation (\ref{eq:best_para_HXD}).
Thus, the consecutive 6 profiles have indeed lined up in their phases.
A small subpeak, just prior to the minimum,
is seen in at least 4 out of the 6 profiles.
These results reinforce the validity of the demodulation analysis.
Figure~\ref{fig:6pulse_profiles}d reproduces,
and directly compares, 
the time-averaged 15--40 keV pulse profiles
before and after the demodulation.
Thus, the pulse amplitude has been nearly doubled.

\subsubsection{Significance of the phase modulation}
\label{subsec:significance}

Although the phase demodulation procedure 
largely increased the HXD pulse amplitude 
(and hence the pulse significance),
we need to examine
whether this increase is statistically significant,
or merely  due to a chance superposition
of  Poisson noise of the data
on the already existing pulse periodicity at $P_{\rm 0h}$.

The nominal chance probability of finding values of  
$X \geq X_{\rm max}$ of equation (\ref{eq:Z2_max}) is extremely small, 
${\cal P}_{2}^{\rm h} = 3.0 \times 10^{-9}$.
However, we cannot use this face value 
to evaluate the statistical significance of the detected effect,
for the following two reasons.
\begin{enumerate}
\item This ${\cal P}_{2}^{\rm h}$ must be multiplied by the total number of 
{\it independent}  trials 
involved in the computation of  figure~\ref{fig:demodulation_HXD},
but this quantity is not easy to estimate.
The total number of parameter search steps,
actually employed,
would considerably overestimate this number, 
because the steps were chosen, rather arbitrarily,
to be fine enough not to miss the $X$ peaks.

\item 
When evaluating the significance of $X_{\rm max}$,
we need to consider the fact 
that the HXD signals were already pulsing,
before the demodulation procedure,
at $P_{\rm 0h}$ (Paper I) 
with a significance of $X_0 \gg n$.
The variations in the pulse amplitude, profiles, and phase
(figure~\ref{fig:6pulse_profiles}) would further contribute to $X$.
\end{enumerate}
These two problems can be combined into a single question:
when we scan over $P$, $A$, $\phi$ and $T$,
like in figure~\ref{fig:demodulation_HXD},
what is the overall chance probability for the $X$ statistics 
to increase {\it by chance} from $X_0$  to $X_{\rm max}$ or higher 
when the source is pulsating with variable parameters
but without any excess variation at $T$ above the white noise?

To answer the above  inquiry,
we conducted 1,000 simulation runs
as detailed in Appendix,
using the actual 15--40 keV data themselves,
but purposely randomizing, 
each time, the phase of the assumed 36 ks modulation:
any white noise
would not be affected by such randomization.
Figure~\ref{fig:chance_probability}a shows a distribution 
of the grand-maximum value $X_{\rm max}$ obtained in this simulation study.
It is considered to represent the probability distribution of $X_{\rm max}$,
to be found in the same demodulation analysis of the HXD data
over the same  $(P, T, A, \phi)$ space as figure~\ref{fig:demodulation_HXD},
but under the absence of  
power enhancement (above a white noise) at any  $T$ examined.
There, values exceeding  equation (\ref{eq:Z2_max})
were actually found in three out of the 1,000 runs,
namely, with a probability of 0.3\%.

Figure~\ref{fig:chance_probability}b gives 
the upper integral probability associated with figure~\ref{fig:chance_probability}a. 
In the range of $X_{\rm max}>35$, it is approximated  by an exponential as
\begin{equation}
f(>X_{\rm max}) = f_0 \exp \left(-\frac{X_{\rm max}-40}{g_0} \right)
\label{eq:chance}
\end{equation}
with $f_0= 0.038 \pm 0.005$
and $g_0 = 2.2 \pm 0.2$,
where $X_{\rm max}=40.0$ was chosen as a reference pivot point.
The value of $g_0$ is consistent with the exponential factor, 2.0,
that is expected for a chi-square distribution of any degree of freedom.
Thus, a value of $X_{\rm max}$ 
which is equal to or larger than equation (\ref{eq:Z2_max}) 
will appear with a chance probability of 
\begin{equation}
f(45.60) = (3 \pm 1) \times 10^{-3}~,
\end{equation}
in a good agreement with the implication of panel (a).
As noticed in Appendix, this estimate is considered rather conservative.

Considering that MEA14 used $n=4$,
it may be interesting to examine what happens if  different values of $n$ are employed.
We hence  repeated the same analysis with $n=1$, 3, and 4,
to find $X_{\rm max} = 25.64$, 47.06, and 49.92, respectively.
The optimum parameters were not much different  
from those in equation (\ref{eq:best_para_HXD}).
Although the larger harmonic numbers somewhat 
increased $X_{\rm max}$ from equation (\ref{eq:Z2_max}),
the effect is not significant considering the increased degree of freedom.
In fact, the implied chance occurrence probability, 
${\cal P}_{1}^{\rm h} =  2.8 \times 10^{-6}$,
${\cal P}_{3}^{\rm h} =  1.8 \times 10^{-8}$,
and ${\cal P}_{4}^{\rm h} =  4.2 \times 10^{-8}$,
are all much higher than ${\cal P}_{2}^{\rm h}$ obtained above,
in agreement with the double-peaked pulse profiles.
Therefore, we retain $n=2$ below.

In short, we can exclude, at $>99.6\%$ confidence, the possibility
that  $X_{\rm max}$ of equation (\ref{eq:Z2_max})
in figure~\ref{fig:demodulation_HXD} arised
via chance superposition of the photon count fluctuations
on top of the hard X-ray pulsation 
of which the  parameters vary with white noise.
This confidence level is more stringent than the case of 4U 0142+61,
where it was $>99\%$ (MEA14).

\subsection{Analysis of the XIS data}
\label{subsec:XIS}

The two-component spectral model,
fitted in Paper I to the XIS+HXD spectrum of 1E 1547.0$-$5408,
indicates that the soft thermal component 
and the hard power-law component 
cross over at $\sim 5$ keV.
Therefore, the hardest end of the XIS energy band,
with its P-sum time resolution, is expected
to provide independent information on the hard component,
and allows us to  crosscheck the HXD-PIN results.
(This was not feasible with 4U~0142+61,
since the XIS was not operated in the P-sum mode
in that observation.)
To limit the soft-component contribution to $<10\%$,
we choose an energy range of 10--14 keV,
and perform  the same analysis as above.
Although this energy range,
which is  in fact covered by the XIS,  
is seldom utilized in spectral studies
mainly because of the calibration uncertainty therein,
such problems do not affect timing studies.

Figure~\ref{fig:demodulation_XIS} show the results 
of this demodulation analysis on the 10--14 keV XIS0 data,
presented in the same way as figure~\ref{fig:demodulation_HXD}.
Panels (a), (b), and (c) all refer to the condition 
of $T=36.0$ ks after equation (\ref{eq:best_para_HXD}).
On the $(A, \phi)$ plane, a clear peak with 
\begin{equation}
X_{\rm pk}(36{\rm ks})=27.64
\label{eq:Z2_max_XIS}
\end{equation}
is seen at  a similar position to figure~\ref{fig:demodulation_HXD}a.
More specifically, $A$ and $\phi$ can be constrained as 
\begin{equation}
A=0.46 \pm 0.11~{\rm s}, ~~ \phi= 260^{\circ} \pm 17^{\circ}~.
\label{eq:best_para_XIS}
\end{equation}
These values agree, within errors, 
with the corresponding results in equation (\ref{eq:best_para_HXD}).
If the modulation parameters are fixed to the best-fit values of
equation (\ref{eq:best_para_HXD}) as determined with the HXD,
we instead find
\begin{equation}
X_{\rm pk}(36{\rm ks})=17.08~.
\label{eq:Z2_max2_XIS}
\end{equation}

The demodulation with equation (\ref{eq:best_para_XIS})
has effects on the 10--14 keV periodograms 
as shown in figure~\ref{fig:PG_Pr_XIS}a.
While the value of $X\sim 5$ found before the demodulation (black)
is within random fluctuations, a clear peak,  
of which the height is given by equation (\ref{eq:Z2_max_XIS}),
appeared at a period that is in full agreement with 
equation (\ref{eq:P0_XIS}) and equation (\ref{eq:P0_HXD}).

Figure~\ref{fig:PG_Pr_XIS}b shows 10--14 keV 
XIS0 pulse profiles folded at $P_{\rm 0s}$,
obtained using the demodulation parameters
of equation (\ref{eq:best_para_XIS}) 
and presented in the same colors as in figure~\ref{fig:6pulse_profiles}d.
Again, the demodulation procedure significantly increased the pulse fraction.
The red pulse profile is similar to
that in figure~\ref{fig:6pulse_profiles}d,
although the relative depths of the two pulse minima are reversed.
For reference, 1--10 keV XIS0 pulse profile is also shown.

The $T$-dependence of  $X_{\rm pk}$
is shown in figure~\ref{fig:demodulation_XIS}d.
Thus, in addition to a small  hump at $T=36-38$ ks,
we observe many other peaks, some being much higher.
Among them,  
the sharp feature at $T\sim 11$ ks
is  likely to be caused by background variations,
because it is close to $2 P_{\rm orb}$.
Besides, the plot reveals a prominent peak at $T=44$ ks,
which is probably due to a chance superposition of
Poisson fluctuations on top of the demodulated pulsation.
As a result, we cannot claim the presence of the $T=36$ ks
periodicity based on the XIS data alone.
Nevertheless, the XIS results significantly reinforce
the HXD-detected phase modulation:
since the pulsation is insignificant in the raw periodogram
(figure~\ref{fig:PG_Pr_XIS}a),
 the probability of finding a peak of 
$X_{\rm pk}= 17.08$ of  equation (\ref{eq:Z2_max2_XIS})
at the same $(P, T, A, \phi)$ point
as specified by the HXD result,
can be directly evaluated  with a simple chi-square distribution to be  
${\cal P}^{\rm s}_{2}= 1.9 \times 10^{-3}$.
If we instead use equation (\ref{eq:Z2_max_XIS}),
the probability further decreases to 
${\cal P}^{\rm s}_{2}= 1.5 \times 10^{-5}$.

\subsection{Energy dependence of the modulation amplitude}

Now that the 36 ks phase modulation in the hard X-ray pulses,
first detected with the HXD, 
has been reinforced by the 10--14 keV XIS data,
our next  task is to examine 
how this phenomenon behaves towards lower energies.
In the same way as above,
we hence analyzed the XIS0 data in three lower energies;
7--10 keV where the spectral hard component is still dominant,
4--7 keV where they compete,
and 1--4 keV dominated by the soft component.
Figure~\ref{fig:demodulation_softX} shows the results 
of this soft X-ray analysis, namely, scans in $A$
in which $T=36.0$ ks is fixed while  $\phi$ is allowed to take any value.
Thus, the 36 ks modulation amplitude cleary decreases towards lower energies, 
and  the effect almost disappears in the 1--4 keV range.

Taking the 7--10 keV case as an example,
we again calculated the 12-bin folded pulse profiles  for 6 phases of $T=36$ ks,
in a similar manner to figure~\ref{fig:6pulse_profiles}b.
The result, shown in figure~\ref{fig:demodulation_softX}d,
again visualizes the sinusoidal phase modulation,
 clearly with a smaller amplitude than in figure~\ref{fig:6pulse_profiles}b.
The amplitude can be read as  
$\Delta k/K  \sim \pm 2/12= \pm 0.17$ pulse phase, or  $A=0.17 P_{\rm 0h} =0.35$ s, 
in a good agreement with figure~\ref{fig:demodulation_softX}a.
In this case, the pulse peak is more clearly traced than the valley.
Furthermore, the modulation phase is somewhat shifted
from  equation (\ref{eq:best_para_HXD}) to $\phi \sim 360^\circ$.

Combining  figure~\ref{fig:demodulation_softX} with
equation (\ref{eq:best_para_HXD}) and equation (\ref{eq:best_para_XIS}),
the energy dependence of $A$ is  visualized in figure~\ref{fig:amp_vs_energy}.
There, the dashed curve represents 
the fractional contribution of the hard component at that energy, namely
\begin{equation}
A = A_0 \times F_{\rm h}/(F_{\rm s}+F_{\rm h})
\label{eq:amp_vs_energy}
\end{equation}
where $A_0=0.5$ s is a fiducial amplitude at  the hardest energy limit,
while $F_{\rm s}$ and $F_{\rm h}$ are 
the intensities of the soft and hard components, respectively,
according to the spectral decomposition (Model B) in Paper I.
Interesting enough,
this empirical quantity behaves very similarly to 
the observed energy dependence of $A$,
suggesting that the phase modulation is limited to the hard X-ray component.

Given figure~\ref{fig:amp_vs_energy},
we attempted to divide the 15--40 keV HXD-PIN band into finer ranges,
or to extend the analysis to higher energies beyond 40 keV
up to $\sim 70$ keV.
However, due to limited statistics,
the data provided no significant information 
on the possible energy dependence of the HXD phase modulation.
Likewise,  the  HXD-PIN data below 15 keV 
(down to ~10 keV)  had so poor statistics 
that they did not allow us to cross confirm the 10--14 keV XIS results.
All what can be said is
that the pulse-phase modulation in the HXD-PIN signal is 
consistent with being energy independent 
over the 10--70 keV range within rather large uncertainties.

\subsection{Analysis of the HXD-GSO data}
\label{sec:GSO}

As described in Paper I,
the signal from 1E 1547.0$-$5408 was marginally
detected in the 60--100 keV range with HXD-GSO.
Therefore, we analyzed the HXD-GSO data in various energy ranges from 50 to 200 keV,
but the data  gave significant evidence  for neither the pulsation,
nor its phase modulation.
This is not surprising, because HXD-GSO has
a considerably higher background than HXD-PIN \citep{HXD2}.

Even though the HXD-GSO data provide no information
on the source pulsation,
they allow us to perform  an important control study.
We  hence purposely chose a very high energy range, 300--307 keV,
which is dominated by background
and has nearly the same counting rate, 0.46 ct s$^{-1}$, 
as the background-inclusive 15--40 keV HXD-PIN rate
(0.44 ct s$^{-1}$; figure~\ref{fig:6pulse_profiles}).
The green trace in figure~\ref{fig:demodulation_HXD}d
shows the derived $X_{\rm pk}$ behavior in this high energy band.
Thus, the baseline value of $X_{\rm pk} \sim  10$ is 
much lower than that in 15--40 keV,  $X_{\rm pk} \sim  22$,
due to the absence of the pulsation.
In addition, no peaks are observed around $T \sim 36$ ks.
Since these HXD-GSO data were acquired by the same instrument
as the HXD-PIN data,
at completely the same time,
and under the same background conditions,
the result argues against instrumental origins of the 
$T=36$ ks periodicity in the 15--40 keV pulse phases.

\section{Discussion}
\label{sec:discussion}

Through an extensive reanalysis of the Suzaku data of 1E 1547.0$-$5408
acquired in the 2009 outburst,
we obtained evidence that the 2.07 s hard X-ray pulsation
detected with the HXD is subject to 
phase modulation with a period of $T=36$ ks
(figure~\ref{fig:demodulation_HXD}).
Following the first detection from 4U 0142+61 
($P_0 = 8.69$ s, $T=55$ ks, and $A\sim 0.75$ s; MEA14),
this makes 1E 1547.0$-$5408 a 2nd magnetar 
exhibiting  this kind of evidence. 

\subsection{Trivial possibilities}
\subsubsection{Instrumental origin}
The detected effect should be first examined
for instrumental or observational artifacts.
In section \ref{subsec:significance}, we excluded
the statistical fluctuation origin at a 99.6\% confidence level.
Furthermore, the HXD-GSO data analysis (section \ref{sec:GSO}) indicates
that the HXD instrument does not suffer unexpected artifacts
(e.g., background variations)
that can explain away the $X_{\rm pk}$ maximum at $T=36$ ks.
The value of $T=36.0$ ks $= 6.4 P_{\rm orb}$
is not in harmonic resonance with the Suzaku's orbital period, either.

In this respect,
of particular importance is the cross confirmation of the phase modulation
with the 10--14 keV XIS data (figure~\ref{fig:demodulation_XIS}).
This further rules out the possibility 
of  the effect being due to some artifact specific to the HXD.
Furthermore, the clear decrease of $A$ towards lower energies
(figure~\ref{fig:amp_vs_energy}) excludes any
instrumental or artificial origin specific to the overall Suzaku mission
(e.g., unexpectedly large drifts in the spacecraft clock),
because such effects should be independent of the photon energy.
We are hence left with celestial interpretations.

\subsubsection{Period derivatives}

Based on figure~\ref{fig:6pulse_profiles}b,
figure~\ref{fig:5ltcvs}a, and figure~\ref{fig:5ltcvs}c,
we have so far regarded the phase modulation as periodic,
and performed the demodulation analysis
to arrive at the modulation period of equation (\ref{eq:best_para_HXD}).
However, the total data span, 86 ks, 
contains only 2.4 cycles of the 36 ks oscillation.
Hence, we could be mistaking some other effects
for the periodic phase changes.

One possibility is the period derivative, because
magnetars generally spin down,
presumably emitting magnetic dipole radiation. 
This will cause parabolic changes in the pulse phase,
which could mimic a periodic phase modulation.
From Entry No.3 of Table 4 in \citet{Kuiper12},
the average spin-down rate of 1E 1547.0$-$5408 is considered 
to  be $\dot{P}\sim 2 \times 10^{-11}$s s$^{-1}$ 
around the present observation.
Although this would cause a phase difference of only $\sim 0.1$ pulse cycle
across the 86 ks length of the present data,
there still remains a possibility 
that the instantaneous $\dot{P}$ was considerably higher.

With the above in mind,
the $Z_2^2$ analysis was repeated on the 15--40 keV HXD-PIN data,
incorporating this time the $\dot{P}$ term.
We defined the period $P_{\rm 0}$ 
as the value at the middle point of the observation,
and scanned it over the same joint error range as
defined by equation (\ref{eq:P0_XIS}) and equation (\ref{eq:P0_HXD}).
The analysis was first conducted assuming no phase modulation ($A=0$).
However, the pulse significance did not increase significantly;
instead, the $X$ value decreased from equation (\ref{eq:Z2_0})
as $|\dot{P}|$ is increased.
Thus,  a loose limit was obtained as
$-1.5\times 10^{-10} < \dot{P} < 3.2 \times 10^{-10}$ s s$^{-1}$.
We next fixed $T=36$ ks,  and examined the parameter region
around equation (\ref{eq:best_para_HXD}).
However, the pulse significance did not increase, either.
The obtained limit is again uninteresting,
$-1.1\times 10^{-10} < \dot{P} < 0.8 \times 10^{-10}$ s s$^{-1}$.
In short, period derivatives cannot explain the
observed pulse-phase modulation effects.

\subsubsection{Non-periodic origin}
Another possibility is that the hard X-ray pulse profile (including its phase)
is fluctuating randomly for some unspecified reasons,
and  the variability power happened to be high at $T \sim 36$ ks.
To help examination of this case,
the wave number $q\equiv 86{~\rm ks}/T$ is drawn in blue
at the top of figure~\ref{fig:demodulation_HXD}d.
Thus, the significant excess in $X_{\rm pk}\equiv Z_2^2$ 
above  equation (\ref{eq:Z2_0}), e.g., $X_{\rm pk}>30$, 
is confined to a rather narrow range of  the wave number
as $q=2.4 \pm 0.5$,
in a good agreement with figure~\ref{fig:5ltcvs}c
and equation (\ref{eq:best_wavenum}).
Furthermore, we have already confirmed in section \ref{subsec:periodic_modulation}
that the power enhancements at $q=2$ and $q=3$ in figure~\ref{fig:5ltcvs}c
are 3 sigma effects.
These facts altogether mean that the phenomenon 
should be regarded as periodic rather than random.

There is additional evidence arguing against
random phase fluctuations, at least $10< T<30$ ks is concerned.
In this range, the  15--40 keV $X_{\rm pk}$ values  in 
figure~\ref{fig:demodulation_HXD}d
are comparable to $X_0$,
and are all found at $A<0.15$ s,
implying that  the pulse significance {\it diminishes} 
by demodulation with $T$ in this range.
Over a longer period range of $30< T <50$ ks,
$X_{\rm pk}$ in figure~\ref{fig:demodulation_HXD}d
is obtained at $A=0.5-0.6$ s,
as a simple continuation from equation (\ref{eq:best_para_HXD}).
At $T>50$ ks,  the values of $A$ become rather poorly constrained
because of the insufficient data length compared to $T$.
Nevertheless, the pulse-phase modulation is not likely
to contain significant power at these long  periods, either,
because the  power spectrum in figure~\ref{fig:5ltcvs}c shown in red
is not enhanced at $q=1$  above the Poissonian contribution.
Therefore, the pulse-phase modulation is 
unlikely to  have ``pink noise" nature
including secular changes,
which would usually contain significant power at $q=1$

To assess the above issue yet from another aspect,
we  applied the demodulation analysis separately 
to the first and second halves of the 15--40 keV HXD data.
The procedure is the same as in subsection~\ref{subsec:demod_HXD},
except that $T=36.0$ ks was kept fixed because of the shorter data length.
Then, the first half gave the significance peak with $X_{\rm max} \equiv X_{\rm pk}(36)=32.4$
at $A=0.58 \pm 0.18$ s and $\phi=272^\circ \pm 18^\circ$, 
while the second half yielded $X_{\rm max}=22.7$
at $A=0.42 \pm 0.17$ s and $\phi=293^\circ \pm 30^\circ$.
In terms of the parameter $(A,\phi)$, 
the two halves agree within errors with each other,
and with equation (\ref{eq:best_para_HXD}).
Thus, the 36 ks pulse-phase modulation is considered
to have persisted through the present observation.

Finally,  figure~\ref{fig:Pr_1st_2nd}  shows
the 15--40 keV pulse profiles in the 1st and 2nd halves of the data,
demodulated using the common set of parameters
given by equation (\ref{eq:best_para_HXD}).
In fact, we discarded the initial 14 ks of the 1st half
and the final 14 ks of the 2nd half,
to make them both represent exactly the time period of 36 ks.
Within the photon count fluctuations,
the two profile agree with each other,
with $\chi^2=0.78$ for $\nu=12$
(calculated before applying the running average).
Thus,  the pulse profiles from the 1st and 2nd modulation cycles,
when demodulated using the common set of parameters,
become consistent with each other.

In conclusion, we can regard  the phase variations
in the hard X-ray pulses as periodic,
in spite of the rather limited data span.

\subsection{Interpretation as free precession}
\subsubsection{Free precession}
\label{subsubsec:free_precession}
Let us hereafter limit our interpretation to {\it periodic} celestial effects.
The simplest of them is binary motion of the magnetar:
indeed, $A=0.5$ s and $T=36$ ks can be explained
if the putative companion star has a mass
of the order of 0.1 solar mass (depending on the inclination).
However, even putting aside the astrophysical reality of such a binary,
the absence of the phase modulation in soft X-rays
(figure~\ref{fig:amp_vs_energy}) clearly rules out this interpretation, 
just as in 4U~0142+61.

From the above arguments,
free precession  (\cite{Treves,Heyl02}; MEA14) 
described in section~\ref{sec:intro} 
remains the most promising possibility.
In this scenario, we identify  $\hat{x}_3$ with the dipole magnetic axis,
assume that  it is tilted from $\vec{L}$ (i.e., $\alpha \ne 0$),
and consider that the NS is  deformed to have 
a non-zero value of $\epsilon = \Delta I/I$.
Then, $P_{\rm rot}$ and  $P_{\rm prec}$,
which were degenerate when $\epsilon=0$,
split into slightly different two values
according to equation (\ref{eq:epsilon}).
Of them,  $P_{\rm prec}$ around $\vec{L}$ is 
observed as the pulse period of equation (\ref{eq:P0_XIS})
and equation (\ref{eq:P0_HXD}), i.e., $P_{\rm prec}=P_0$, 
while $P_{\rm rot}$ around $\hat{x}_3$ cannot be detected
when the emission pattern is symmetric around  $\hat{x}_3$
even if $\epsilon \ne 0$ and $\alpha \ne 0$ both hold.
This case is thought to apply to the soft component,
because it is presumably thermal emission from the NS surface,
and should be symmetric around the magnetic axis along which the heat will flow.

If, in contrast, the emission pattern breaks symmetry around $\hat{x}_3$,
the observed signals would depend 
not only on the direction of $\hat{x}_3$ as seen from us,
but also on the NS's rotation angle around $\hat{x}_3$
relative to the $\vec{L}$-$\hat{x}_3$ plane.
Since this relative angle changes with the slip period $T$ of equation (\ref{eq:slip}),
the  pulse phase becomes  modulated at $T$  (MEA14).
With a natural identification of  $T$ with the modulation period of 36 ks,
this case can explain the behavior of the hard X-ray component,
and hence equation (\ref{eq:slip}) yields
\begin{equation}
|\epsilon| = |\Delta I/I |= P_0/T = \left(0.58^{+0.04}_{-0.07} \right)\times 10^{-4}~.
\label{eq:epsilon_1547}
\end{equation}

\subsubsection{Double periodicity}
\label{subsubsec:double_periodicity}

If the above interpretation is correct,
we should be able to observe periodicity also at
$P_{\rm rot} = P_{\rm prec}/(1+\epsilon)$.
From equation (\ref{eq:epsilon_1547}),
it should appear as a {\it prograde} period at 
\begin{equation}
P_{\rm rot}^{+} =2,072.027\pm 0.014~{\rm ms}
\label{eq:prograde}
\end{equation}
if the deformation is prolate ($\epsilon>0$),
and a {\it retrograde} period at
\begin{equation}
P_{\rm rot}^{-} =2,072.267\pm 0.014~{\rm ms}
\label{eq:retrograde}
\end{equation}
if oblate ($\epsilon<0$),
although they could appear simultaneously.
Of them,  $P_{\rm rot}^{+}$ is 
already visible in figure~\ref{fig:periodogram0}b.
Nevertheless, the issue deserves further examination,
because the periodicity at  $P_{\rm rot}$ would also 
be subject to the phase modulation at $T$,
and the significance would change by the demodulation procedure.

Figure \ref{fig:PG_superposed} shows all the periodograms 
obtained while computing figure~\ref{fig:demodulation_HXD},
superposed onto a single plot.
In addition to the main peak coincident with equation (\ref{eq:P0_HXD}),
the results thus reveal clear enhancements at
$\sim 2,072.03$ ms and $\sim 2,072.25$ ms.
Since they can be readily identified with $P_{\rm rot}^{+}$ and $P_{\rm rot}^{-}$, respectively,
our expectation has been confirmed.
Furthermore, the $P_{\rm rot}^{+}$ peak is somewhat more significant,
in agreement with our assumption that the object has a prolate form.
To more quantitatively asses  the significance of these two enhancements
and their differences remains our future study.

\subsection{Comparison with 4U 0142+41}
In terms of free precession of an axisymmetric NS,
let us compare 1E 1547.0$-$5408 and 4U 0142+41 (MEA14).
The most important feature common to them is 
the similar degree of deformation, 
$\epsilon= 1.6\times 10^{-4}$ in 4U 0142+41 
and $\epsilon=0.6 \times 10^{-4}$ 1E 1547.0$-$5408
[equation~(\ref{eq:epsilon_1547})].
This suggests that the deformation in the two objects has a common origin.
Another resemblance between them is 
the decrease of the pulse-phase modulation amplitude towards lower energies;
at least in 1E 1547.0$-$5408, 
the effect is consistent with being absent in the soft spectral component.
As already described in section~\ref{subsubsec:free_precession},
these properties can be interpreted by considering 
that the soft component is emitted symmetrically around the stellar magnetic axis,
while the hard component breaks that symmetry.
This  difference suggests
that the two spectral components are distinct
not only in their spectral shapes,
but also in their emission patterns and/or emission regions. 

While the two sources thus show similar behavior,
they also exhibit the following three differences,
which must be explained in order to further solidify the interpretation.
\begin{enumerate}
\item  While we measured $A/P_0=0.08 \pm 0.04$ in 4U 0142+61,
the ratio is much higher, $A/P_0=0.25 \pm 0.07$
(approximately $\pm 1/4$pulse cycle)  in 
1E 1547.0$-$4508 in energies above 10 keV
(figure~\ref{fig:amp_vs_energy}).
\item While $n=4$ was needed in 4U 0142+61,
$n=2$ is optimum in 1E 1547.0$-$45408
(section \ref{subsec:significance}).
\item 
The hard X-ray pulsation of 4U~0142+61 in the 2009 Suzaku observation
was detected only after the demodulation,
whereas that of 1E 1547.0$-$5408 is significant
even before such  corrections (Paper I).

\end{enumerate}

Let us consider item No.1 first.
In 4U 0142+61, the sinusoidal hard X-ray modulation 
with $A/P_0=0.08$ was  attributed to  positional (or directional) deviations 
of the X-ray emission region (or direction) from the magnetic poles (or from $\hat{x}_3$).
There, the emission pattern was implicitly 
and without particular reason
assumed to be  pencil-beam like.
In contrast, the case of 1E 1547.0$-$5408 
can be better explained  by assuming 
that  the emission reaches us 
(for simplicity, from a single pole)
 in a {\it fan-beam-like} pattern,
as illustrated in figure~\ref{fig:fanbeam}.
The beam will point to us twice per $P_{\rm prec} =2.07$ s,
when the pole appears from the rim of the NS, 
to be called ``dawn",
and disappears behind it,
to be called  ``dusk".
This explains the double-peaked pulse profiles
(figure~\ref{fig:6pulse_profiles}, figure~\ref{fig:PG_Pr_XIS}),
and the $\sim \pm 1/4$ cycle offset of the hard X-ray pulse peaks
from those in the soft X-rays (figure\ref{fig:PG_Pr_XIS}b).

As already described  in section~\ref{sec:intro} and section~\ref{subsubsec:free_precession},
we  further assume
that the fan-beam emission pattern is asymmetric around $\hat{x}_3$.
Then, at a certain time $t_0$ of the 36 ks slip period,
the signal will be  stronger at the ``dusk" pulse phase than at the ``dawn" phase,
as illustrated in figure~\ref{fig:fanbeam}a;
the pulse peak is delayed by $\sim 1/4$ cycle from the average,
just as in the red to orange modulation phases in figure~\ref{fig:6pulse_profiles}b.
At $t \sim t_0+T/4$,
the configuration will become as in figure~\ref{fig:fanbeam}b,
so that the intensity will be similar between ``dawn" and ``dusk";
the pulse-peak phase will occur at about ``noon" phase with zero delay,
as in the yellow profile in figure~\ref{fig:6pulse_profiles}b.
Then, at $t \sim t_0+ T/2$, 
the signal will become strongest at the ``dawn'' pulse phase, 
as shown by figure~\ref{fig:fanbeam}c;
the pulse peak preceds  the average by $\sim P_{\rm prec}/4$,
as represented by the green to blue profiles in figure~\ref{fig:6pulse_profiles}b.
Finally, at $t \sim t_0+3T/4$,
the pulse amplitude becomes low,
because  the emission region is behind the horizon
when the beam points towards us;
this may correspond to the purple profile in figure~\ref{fig:fanbeam}b.
Thus, the fan-beam configuration with a broken axial symmetry
can explain the large modulation amplitude of 1E 1547.0$-$5408
reaching $A \sim T/4$.
Although the actual configuration (e.g., contribution from the other pole)
would not be as simple as this,
more detailed modeling of the emission pattern 
is beyond the scope of the present paper.

How about the remaining items No.2 and No.3?
In the 2009 data of 4U 0142+61,
the hard X-ray pulse profile was richer in fine structures,
consisting of  three peaks separated by a quarter  cycle
(figure 1f of MEA14).
This is considered to be the reason why $n=4$ was needed.
Moreover, such a complex pulse profile of 4U 0142+61 is thought to 
have been strongly smeared out by the precession,
and hence the hard X-ray pulse was difficult to detect
without demodulation at least in the 2009 Suzaku observation.
In contrast, the pulse profiles of the present source are basically double-peaked.
This is considered to make $n=2$  appropriate for the present source, 
and make the profile less affected by the phase modulation.
In addition, the effects in  1E 1547.0$-$5408 can be 
approximated as a sort of pulse-profile change (figure~\ref{fig:fanbeam}),
with the ``dawn" and ``dusk" intensities changing across the slip period,
rather than a continuous phase shift.
This property is thought to be another factor
that has made the pulse easier to detect without demodulation.
Thus, the differences No.2 and No.3 can be explain in a qualitative way.

\subsection{Relations to other neutron stars}
\label{subsec:otherNSs}

So far, a fair number of reports have been made
on possible detections of precession from NSs,
including accretion-powered objects (e.g., \cite{HerX1_09})
and rapidly rotating NSs 
(e.g, \cite{Precession01,Precession03,Precession06}).
However, as discussed in MEA14,
such a behavior in accretion-powered systems
should be regarded as forced precession, 
because of the strong accretion torque. 
The fast-rotating NSs can exhibit free precession, 
because  they must be deformed into oblate shapes  by centrifugal force.
However, the precession would be damped 
on rather short time scales \citep{Cutler02},
because an oblate body attains its minimum energy
at $\alpha=0$ under a constant $\vec{L}$.
Therefore, these reports remain generally rather unconvincing.

In contrast to these ordinary NSs,
magnetars are expected to more ubiquitously exhibit the free precession,
because their strong $B_{\rm t}$ will deform them
into prolate shapes \citep{Ioka01, Cutler02,Ioka+Sasaki04};
their centrifugal deformation (oblate) is negligible, e.g., $|\epsilon| <10^{-6}$,
because of their rather  long pulse periods, $P=2-11$ s.
Actually, the present data may favor the prolate ($\epsilon>0$) case
(subsection \ref{subsubsec:double_periodicity}).
In such a prolate magnetar, 
the free precession will spontaneously develop,
because internal energy dissipation processes,
with $\vec{L}$ conserved, will increase $\alpha$
on reasonable time scales  \citep{Cutler02}.
(As mentioned in MEA14,
decay in $\alpha$ by emission of gravitational waves is negligible.)
Therefore, the two necessary conditions, 
$\epsilon \ne 0$ and $\alpha \ne 0$,
can be regarded as almost inherent to magnetars with strong $B_{\rm t}$.
Actually, the strong pulsations observed from nearly all magnetars
imply $\alpha \ne 0$.

The remaining requirement for the detection of free precession,
i.e., the non-axi-symmetric emission pattern of the hard component,
is considered more conditional rather than intrinsic.
One possibility is that a magnetic pole on $\hat{x}_3$
is magnetically connected to a nearby local multipole,
and this connecting region is responsible for the hard X-ray generation.
Depending on the detailed magnetic configuration,
such a radiation may well have the required anisotropy around,
and positional offsets from,  $\hat{x}_3$.
Since the magnetic configuration of a magnetar may 
change with time \citep{Magnetar}, e.g., across active states,
$A$ can change even though $T$ and $\alpha$ are kept constant.
Actually, \citet{Kuiper12} found significant pulse-profile changes
in 1E 1547.0$-$5408 across the 2009 activity episode.
Similarly, the hard X-ray phase modulation of 4U 0142+61, 
detected in 2009 and reconfirmed in a subsequent observation in 2013 \citep{Max14},
was absent in an earlier Suzaku observation made in 2007 
(\cite{Enoto11}; see also MEA14).
It is hence of significant interest to clarify
whether the phase-modulation amplitude $A$ of 1E 1547.0$-$5408  
have been changing along with its long-term decay after the 2009 activity.
Future observations of this object with
NuSTAR and ASTRO-H are very important.

\subsection{Toriodal magnetic fields}
\label{subsec:toroidal}

From equation (\ref{eq:slip}) and equation (\ref{eq:epsilon_1547}),
1E 1547.0$-$5408 is concluded to be aspheric 
by $\epsilon =P_0/T=0.6 \times 10^{-4}$,
which is of the same order as that of 4U 0142+61,
$1.6 \times 10^{-4}$ (MEA14).
Then, adopting the magnetic deformation interpretation,
these NS are both inferred via equation~(\ref{eq:epsilon})
to have $B_{\rm t}\sim 10^{16}$ G.
The present result thus gives a strong support 
to the suggestion by MEA14,
that magnetars have ultra-intense $B_{\rm t}$
and are hence deformed to a detectable level.

While 4U~0142+61 is a rather aged magnetar
with a characteristic age of $\tau_{\rm c}=68$ kyr
and $B_{\rm d}=1.3\times 10^{14}$ G,
1E 1547.0$-$5408 has 
$\tau_{\rm c}=1.4$ kyr 
and $B_{\rm d}=2.2\times10^{14}$ G.
Even though $\tau_{\rm c}$ of magnetars can be
systematically overestimated \citep{Nakano15},
1E 1547.0$-$5408 is clearly a much younger 
and more active object than the other,
as evidenced by its faster rotation and higher activity.
The somewhat smaller $\epsilon$ of 1E 1547.0$-$5408 
apparently contradicts to these facts,
and the prospect that magnetars must be
consuming their toroidal (or internal),
as well as dipole (or external), field energies.
Possibly, the $B_{\rm t}$ vs $B_{\rm d}$ ratio may scatter
considerably among mangetars. 
To answer this issue,
we clearly need to increase the number of 
detections of the free precession from magnetars.

\subsection{The wobbling angle $\alpha$}
Finally, we consider possible values of the wobbling angle $\alpha$.
Unlike $\epsilon$ which can be accurately determined via equation (\ref{eq:slip}),
$\alpha$ is rather difficult to unambiguously estimate from the data,
because its effects degenerate with those of the radiation anisotropy.
Below, some qualitative consideration is carried out.

On one hand, $\alpha$ cannot be close to 0,
for the following reasons.
First, an aligned rotator with $\alpha \sim 0$ ($\hat{x}_3 \parallel \vec{L}$)
would not show the soft X-ray pulsation,
as already noticed in section \ref{subsec:otherNSs}.
Second, $\alpha$ would increase
on a reasonable time scale (section \ref{subsec:otherNSs}).
Finally, the dipole field derived from the $\sqrt{P \dot{P}}$ factor
is in reality $B_{\rm d} \sin \alpha$ rather than $B_{\rm d}$ itself,
so $\alpha \sim 0$ would demand too high a value of $B_{\rm d}$.

On the other hand, $\alpha$ cannot be close to $90^\circ$, either.
This is because such an orthogonal rotator would
exhibit a very high pulse fraction
(unless we are observing from a pole-on direction),
contradicting to the fact 
that 1E 1547.0$-$5408 has a particularly low pulse fraction
among the known magnetars.
All what  can be said at present is 
that $\alpha$ is likely to be in an intermediate range
between 0 and $90^\circ$.
A more realistic estimate on $\alpha$ will be available in future
when we model the hard X-ray emission pattern,
and fit its predictions to the actual data.

\section{Conclusoin}
During the 2009 January outburst of 1E 1547$-$5408,
its 2.07 s pulsation exhibited a phase modulation with a period of 36 ks,
in  the 15--40 keV HXD data
as well as in the  XIS data at  energies above 4 keV.
The modulation is consistent with being solely carried by the spectral hard component,
while absent in the soft component.
These results can be understood by presuming
that the NS in this object  is exhibiting torque-free precession
under a deformation by $\epsilon = 0.6 \times 10^{-4}$,
and its hard X-ray emission is asymmetric around the NS's symmetry axis.
These results further suggest the presence of
intense toroidal magnetic fields of $B_{\rm t} \sim 10^{16}$ G.

\section*{Acknowledgements}
The authors are grateful to all the members of the Suzaku team,
for their continued dedication in the spacecraft operation, 
data archiving, and the instrumental calibration.

\section*{Appendix: Phase-randomized demodulation}

The significance of the 36 ks HXD phase modulation 
described in section~\ref{subsec:significance}
was evaluated through a {\it control} study 
using the same actual data themselves, 
rather than via a Monte-Carlo method.
Let us explain how this was carried out.

For this purpose, we again refer to the two-dimensional array $C(j,k)$
(section~\ref{subsec:Pr_evol}),
into which all the 15--40 keV HXD events have been sorted.
The pulse phase and the modulation phase are
represented by $k$ and $j$, respectively.
The $Z_n^2$ calculation effectively produces a folded pulse profile as
\begin{equation}
 D(k) = \sum_{j=0}^{J-1}  C\left(j,k \right)
 \label{eq:Ap1_1}
\end{equation}
and applies the Fourier transform to $D(k)$.
Here, we employ a rather large values of $J$ and $K$,
namely $J=90$ and $K=360$, to avoid binning issues.
The demodulation procedure, instead,
executes the summation by shifting the pulse phase,
according to equation (\ref{eq:modulation}), as
\begin{equation}
 D(k) = \sum_{j=0}^J  C\left(j,k'(j) \right)
  \label{eq:Ap1_2}
\end{equation}
with 
\begin{equation}
 k'(j) = k- K  (A/P_{\rm 0h})  \cdot \sin( 2\pi j/J - \phi )~.
 \label{eq:Ap1_3}
 \end{equation}

Let us assume, as a null hypothesis, 
that  the  pulse phase is intrinsically {\em not} modulated at the period $T$.
Then, the demodulation procedure using
equation (\ref{eq:Ap1_2}) and equation (\ref{eq:Ap1_3}) has two effects; 
(i) to degrade the underlying pulse coherence,
and (ii) to add up Poisson noises of different pixels in various different ways.
Large values of  $X_{\rm max}$,
like in equation (\ref{eq:Z2_max}),
will appear  when  (ii) overwhelms (i).
Importantly, these two effects must remain unchanged
even when a random permeation is applied on the rows $\{j = 0, 1, .., J-1\}$,
because  all rows must be equivalent with respect to (i),
and the Poisson noise in different pixels of $C(j,k)$ 
should be mutually independent when considering (ii).

\begin{figure}[hb]
  \begin{center}
  \FigureFile(70mm,){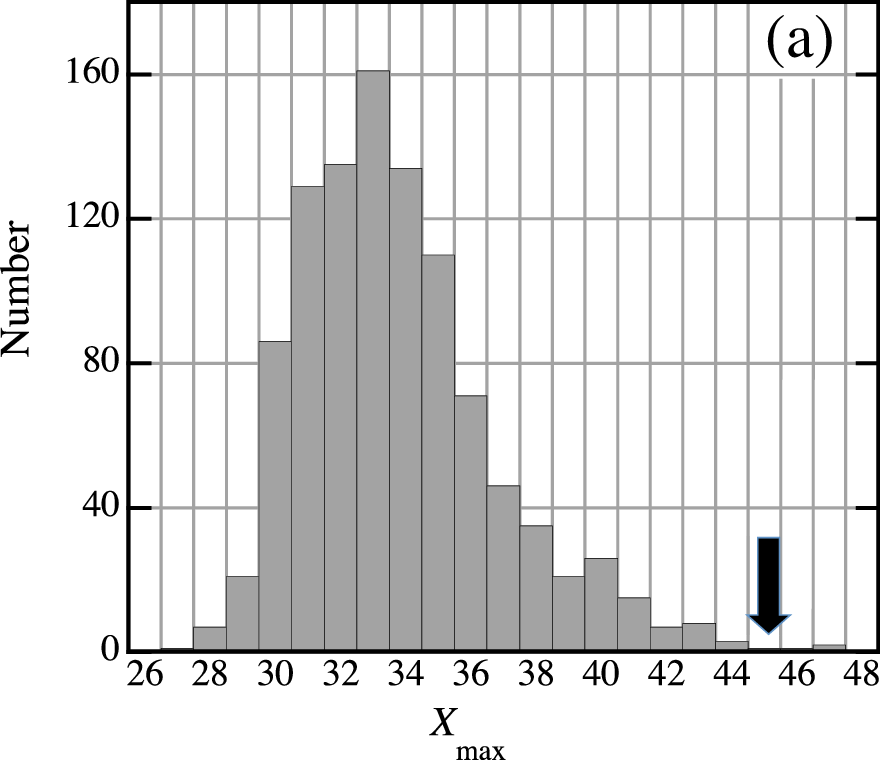}
   \FigureFile(65mm,){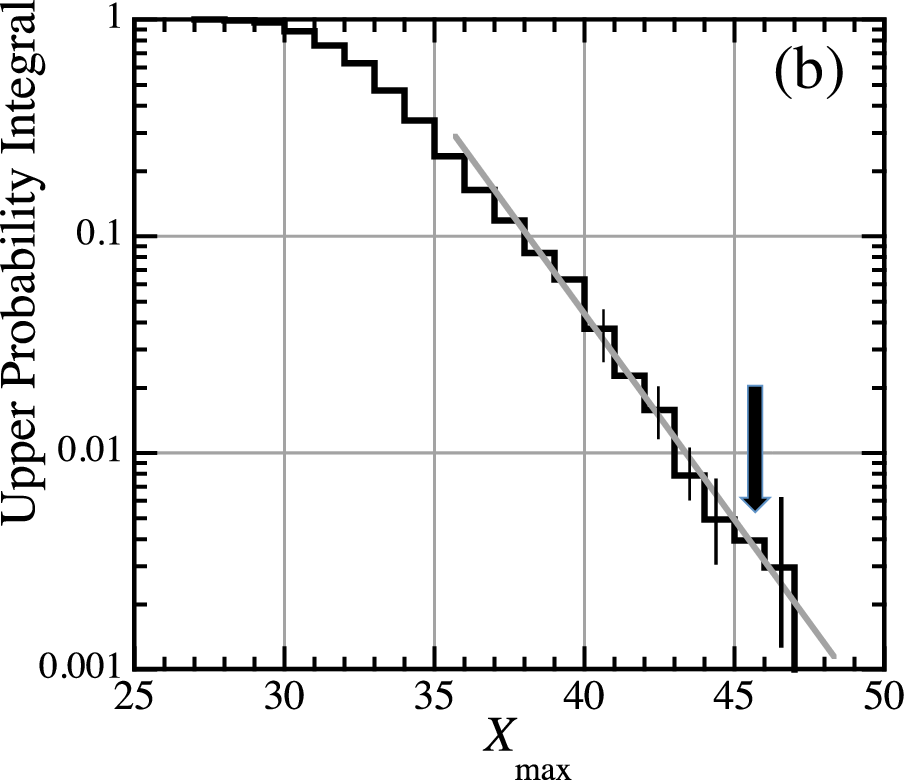}
  \end{center}
  \caption{
A histogram of the values of $X_{\rm max}$ (panel a),
  and the associated upper probability integral (panel b),
  found in a series of 1,000 simulation trials
  each equivalent to figure~\ref{fig:demodulation_HXD}
  but randomizing the 36 ks modulation phase.
  As described in Appendix, it was derived from the data themselves.
   In both panels, a black arrow indicates equation (\ref{eq:Z2_max}).
   }
 \label{fig:chance_probability}
  \end{figure}

\begin{figure}[tbh]
  \begin{center}
   \FigureFile(140mm,){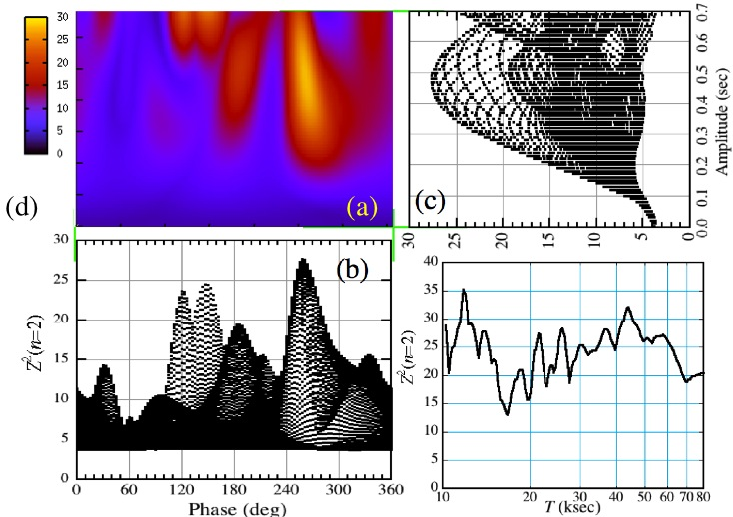}
  \end{center}
  \caption{The same as figure~\ref{fig:demodulation_HXD},
  but using the 10--14 keV XIS0 data.
  }
 \label{fig:demodulation_XIS}
  \end{figure}

\begin{figure}[bth]
  \begin{center}
 \FigureFile(70mm,){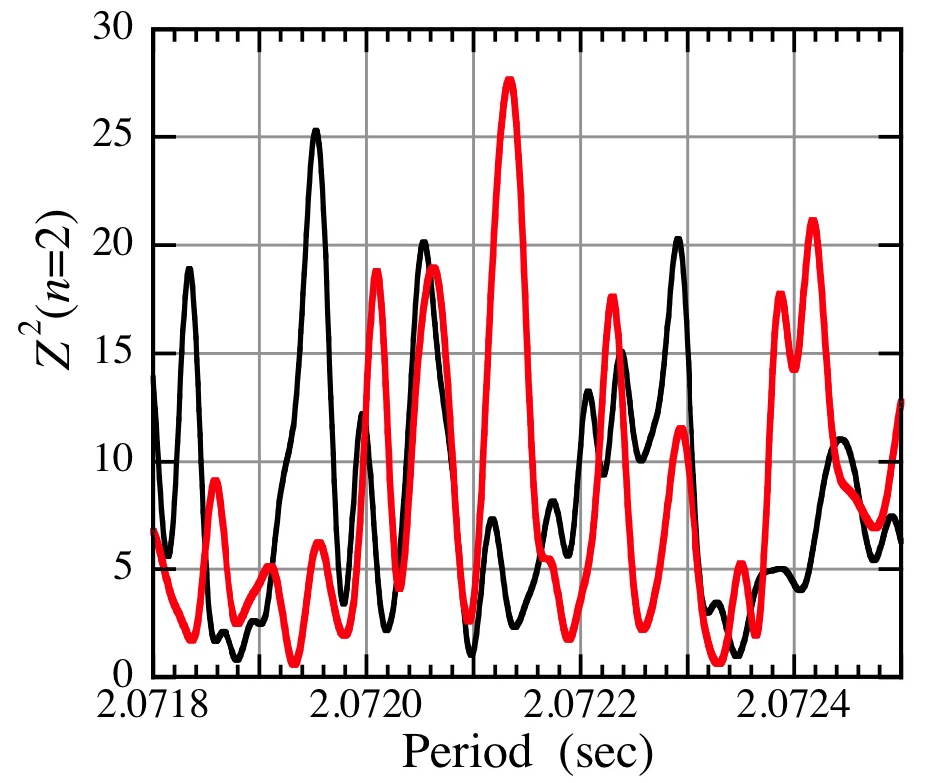}
 \FigureFile(70mm,){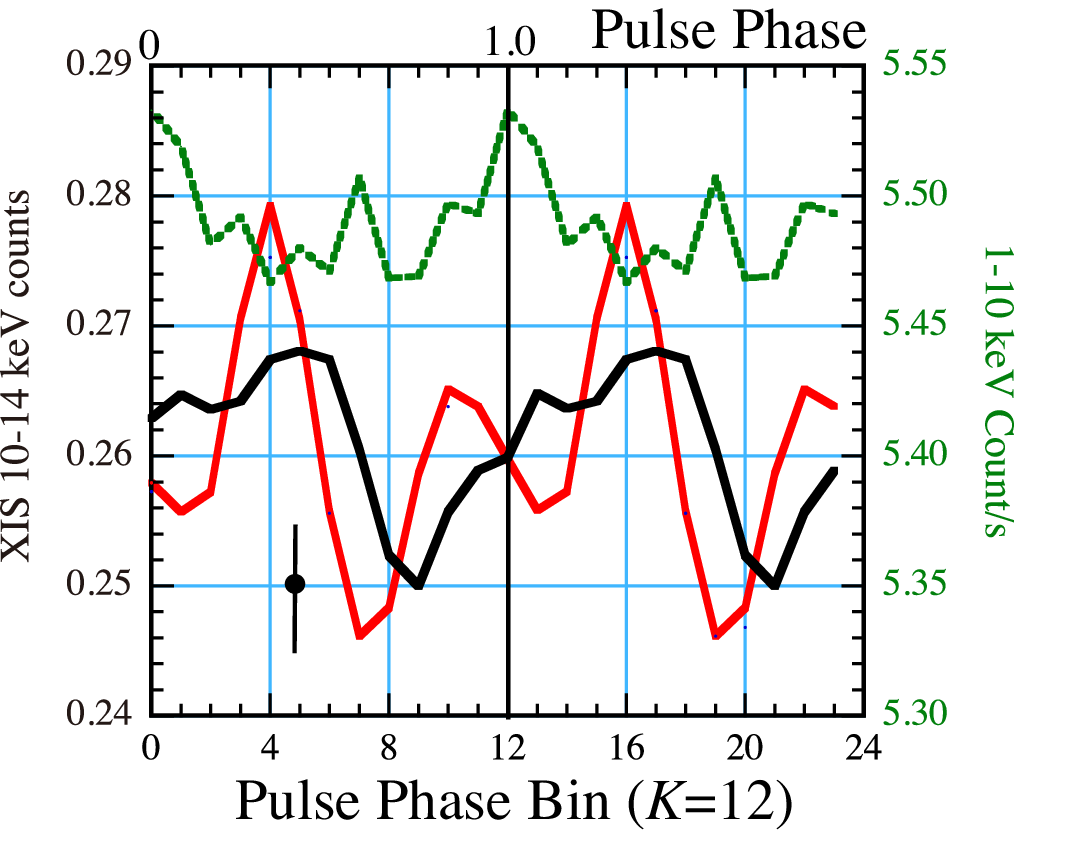}
  \end{center}
  \caption{Effects of the demodulation on the 10--14 keV XIS0 data.
Equation (\ref{eq:best_para_XIS}) and $T=36.0$ ks are employed.
  (a)  Periodgrams, before (black) and after (red) the  demodulation.
   The plot is more expanded in the period range than those presented so far,
  to show finer details around the pulsation.
 (b)  The 10--14 keV pulse profiles folded at $P=P_{\rm 0s}$,
 with the colors  corresponding to those of panel (a).
  Running averages over the adjacent 3 bins are applied.
 The phase origin is the same as in figure~\ref{fig:6pulse_profiles}.
  Green (with right ordinate) shows the 1--10 keV profile.
 }
 \label{fig:PG_Pr_XIS}
  \end{figure}

\begin{figure}[h]
  \begin{center}
 \FigureFile(120mm,){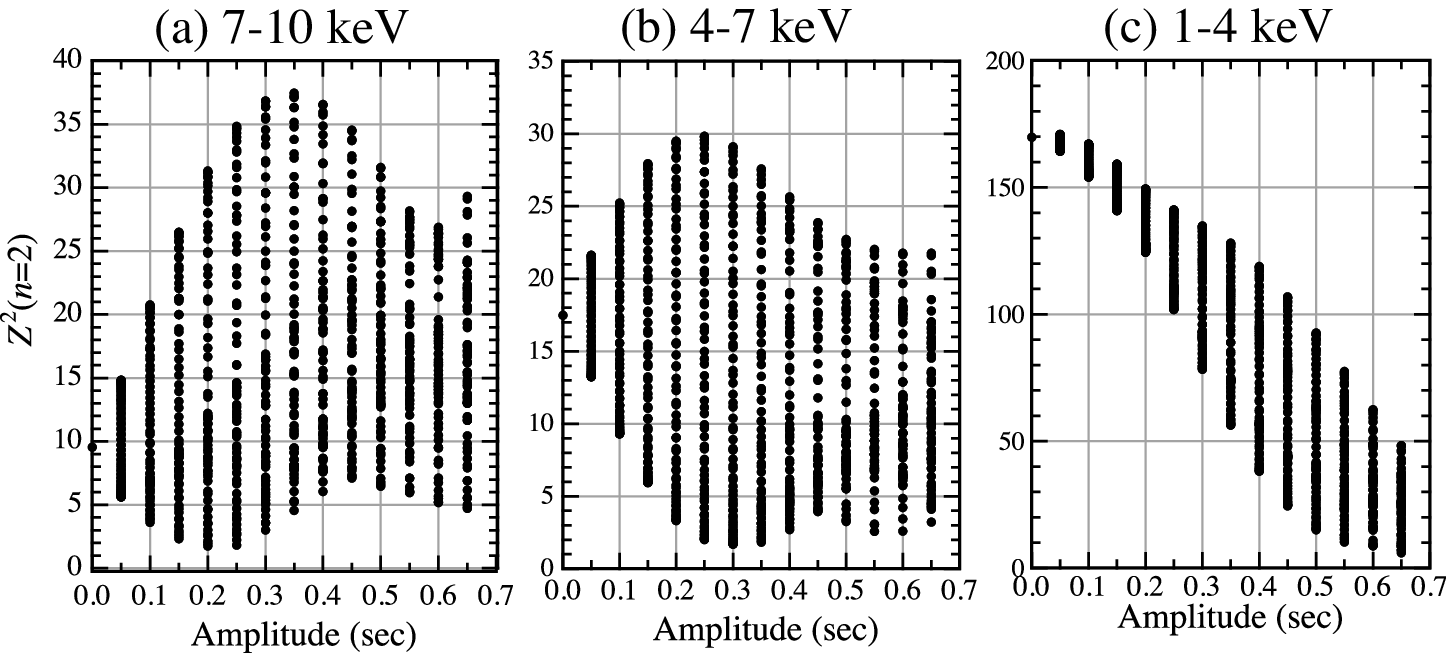} 
 \FigureFile(40mm,){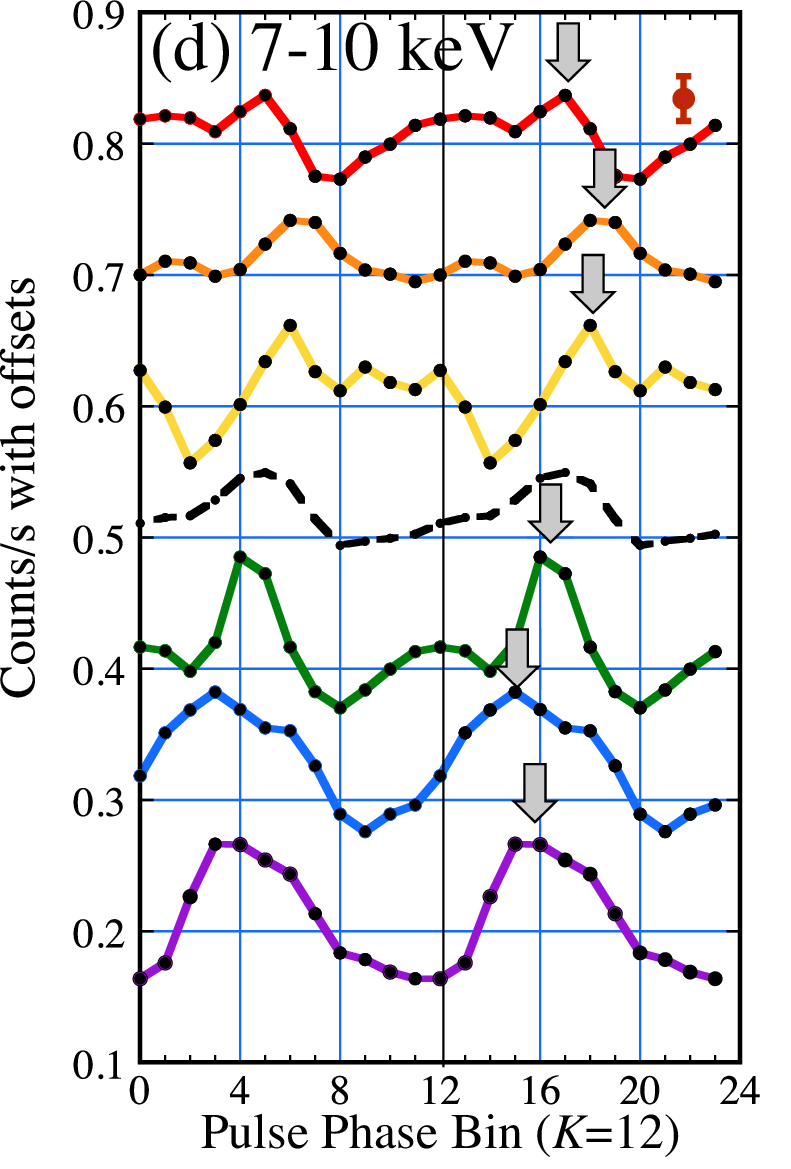} 
  \end{center}
  \caption{
 (a) The same $A$-dependence of $X$ as figure~\ref{fig:demodulation_HXD}c
 and figure~\ref{fig:demodulation_XIS}c,
 but for the 7--10 keV XIS0 data.
  $T=36$ ks is fixed, while $\phi$ is left free.
 (b) In 4--7 keV. (c) In 1--4 keV.
(d) The same as figure~\ref{fig:6pulse_profiles}b but for the 7--10 keV XIS0 data.}
 \label{fig:demodulation_softX}
   \end{figure}

\begin{figure}[tbh]
  \begin{center}
\FigureFile(100mm,){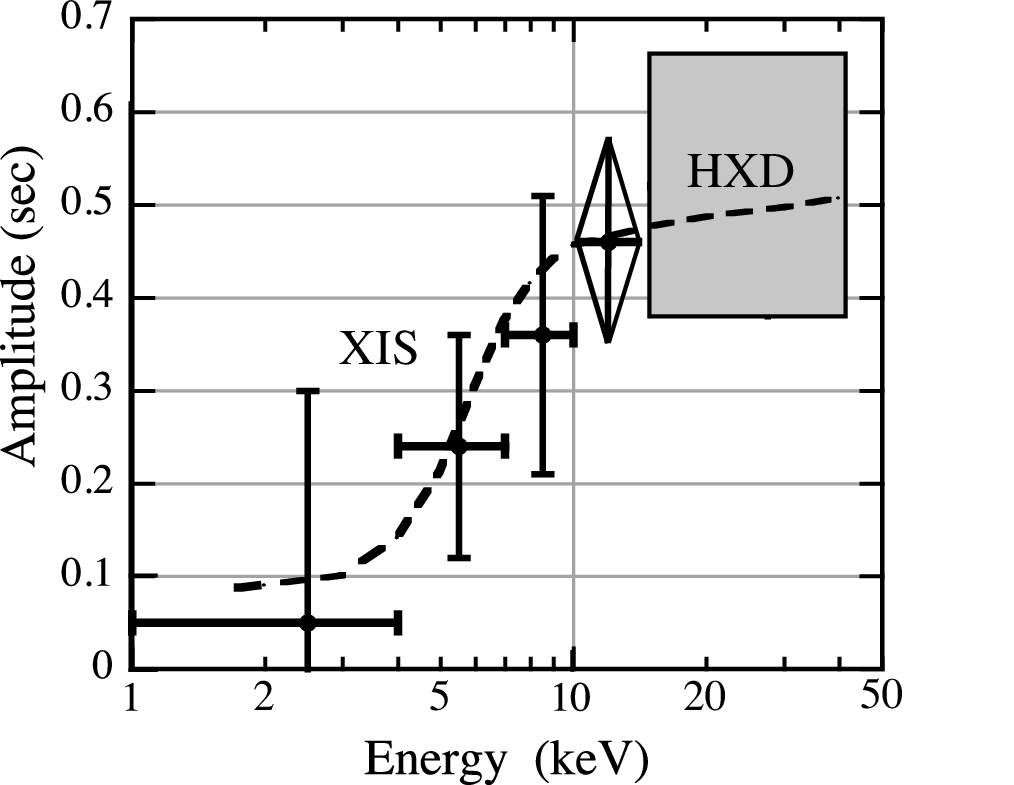}
\end{center}
   \caption{
  The 36 ks phase modulation amplitude $A$,
  obtained at various energies with the HXD and the XIS.
  The modulation phase is not specified.
  The dashed curve represents equation (\ref{eq:amp_vs_energy}).
 }
 \label{fig:amp_vs_energy}
 \end{figure}

\begin{figure}[tbh]
\begin{minipage}{80mm}
  \begin{center}
   \FigureFile(80mm,){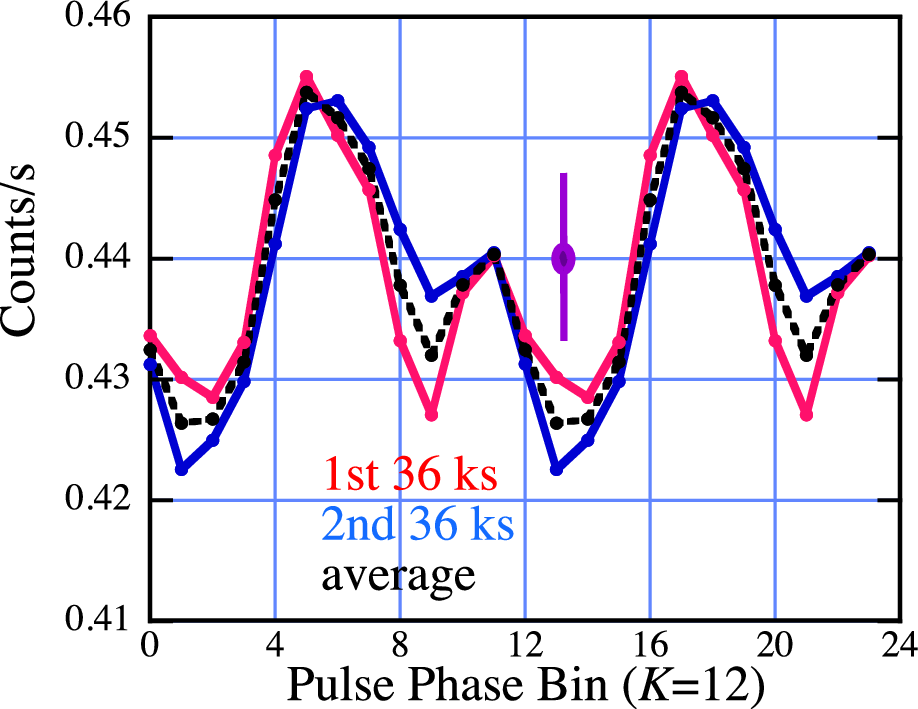}
  \end{center}
\caption{
The 15--40 keV pulse profiles in the 1st (red) and 2nd (blue) 36 ks,
both demodulated using the parameters in equation (\ref{eq:best_para_HXD})
and folded at equation (\ref{eq:P0_HXD}).
Their average, shown in dotted black lines,
can be slightly deferent from the red data in figure~\ref{fig:6pulse_profiles}d
because of the exclusion of some fraction of the data (see text).}
\label{fig:Pr_1st_2nd}
\end{minipage}
\hspace{10mm}
\begin{minipage}{80mm}
  \begin{center}
   \FigureFile(60mm,mm){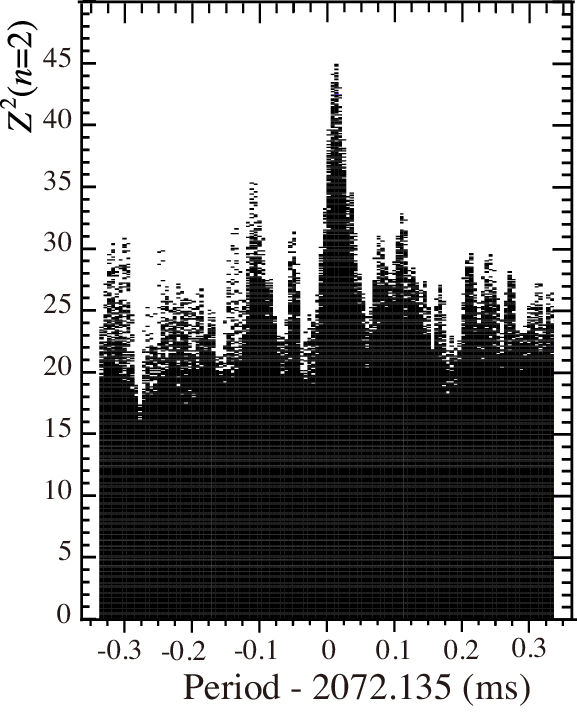}
  \end{center}
\caption{
All the 15--40 keV periodograms  obtained in computing
figure~\ref{fig:demodulation_HXD},
superposed onto a single plot.
The two periodograms in figure~\ref{fig:periodogram0}b
are both included here.
}
\label{fig:PG_superposed}
\end{minipage}
\end{figure}

\begin{figure}[tbh]
  \begin{center}
   \FigureFile(140mm,){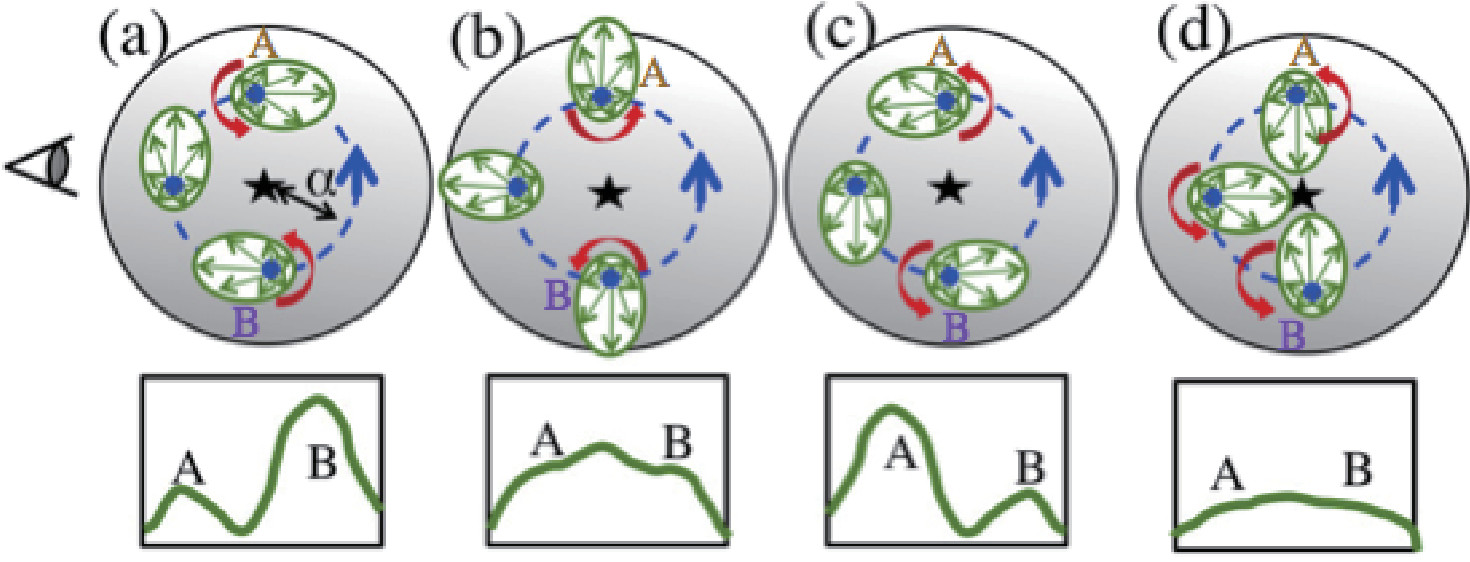}
  \end{center}
  \caption{A possible free precession scenario of an oblate NS,
  as seen from above onto its angular momentum vector $\vec{L}$
  whose tip is represented by a black star.
  The observer is viewing from the left.
  The NS's symmetry pole is represented by a small blue circle,
  and its precession locus by a larger dashed blue circle with a blue arrow.
  Green ellipses show hard X-ray emission pattern,
  which is assumed to be fan-beam like.
  The panels (a), (b), (c), and (d) represent slip-period epochs of 
  $t_0$, $t_0+T/4$,  $t_0+T/2$, and $t_0+3T/4$, respectively.
  A and B indicate ``dawn" and``dusk", respectively.
  Lower panels show the expected hard X-ray pulse profiles. }
 \label{fig:fanbeam}
 \end{figure}

With the above in mind,
we repeated the same demodulation search 1,000 times,
over the same $(P, T, A, \phi)$ volume.
In each trial, $k'(j)$ in the above two equations were replaced with $k'(j')$,
where $\{j' = 0, 1, .., J-1\}$ is a random permutation of $\{j = 0, 1, .., J-1\}$.
Different trials employed different realizations of the permutation,
all with $J=90$.
We thus obtained 1,000 values of $X_{\rm max}$,
of which the histogram is shown in figure~\ref{fig:chance_probability}a.
By integrating this distribution from $X_{\rm max}=+\infty$
towards lower values $X_{\rm max}$,
the upper probability integral has been obtained
as in figure~\ref{fig:chance_probability}b.
This gives the  chance occurrence probability
of large values of $X_{\rm max}$, 
under the presence of the pulsation at $P_0$
but the absence of any intrinsic periodicity at $T$.
Equation~\ref{eq:chance} provides its approximation
at $X_{\rm max} > 35$.

This control study is considered to be rather conservative 
for the following reasons.
When the phase modulation is sinusoidal,
a small difference in the initial value of $\phi$ will not affect $X$,
because $\Delta t$ is expected to have a long coherence length in $\phi$.
This makes a scan over $\phi$ only partially independent.
However, when the random permutation is employed,
$\Delta t$ attains a rather short (typically, $2 \pi/J$ radian) coherence in  $\phi$,
so that scans over $\phi$ become essentially all independent.
As a result, one control run is expected to involve
effectively a lager number of trials.
If this effect is taken into account,
the true chance probability would be 
even lower than equation (\ref{eq:chance}).

\onecolumn

\bigskip

\begin{thebibliography}{99}
\bibitem[Beloborodov et al.(2013)]{Twist07}
 Beloborodov, A. M., \& Thompson, C. 2007, \apj, 657, 967
\bibitem[Braithwaite(2009)]{toroidal09b}
Braithwaite, J. 2009, \mnras, 397, 763
\bibitem[Brazier(1994)]{Zn2_94}
Brazier, K. T. 1994, \mnras, 268, 709
\bibitem[Buccheri et al.(1983)]{Zn2_83} 
 Buccheri, R. et al. 1983, \aap, 128, 245
\bibitem[Butikov(2006)]{Butikov06} 
Butikov, E. 2006,  Europ. J. Phys.,  27, 1071
\bibitem[Carlini \& Treves(1989)]{Treves}
Carlini, A. \& Treves, A. , \aap, 215, 283 (1989).
\bibitem[Chukwude et al.(2003)]{Precession03}
Chukwude, A. E., Ubachukwu, A., \& Okeke, P. 2003, \aap,  399, 231
\bibitem[Cutler(2002)]{Cutler02} 
Cutler, C. 2002, Phys. Rev. D., 66, id 084025 
\bibitem[Dall'Osso et al.(2009)]{toroidal09a} 
Dall'Osso, S., Shore, S.N, \&  Stella, L. 2009, \mnras, 398, 1869
\bibitem[den Hartog et al.(2006)]{denHartog06} 
den Hartog, P. R.,  et al. 2006,  \aap, 451, 587
\bibitem[Enoto et al.(2010b)]{Enoto10b}
Enoto, T., Nakazawa, K.,  Makishima, K., Rea, N.,  Hurley, K., \& Shibata, S. 2010b, 
\apjl, 722, L162
\bibitem[Enoto et al.(2012)]{Enoto12}
Enoto, T., Nakagawa, Y. E., Sakamoto, T., \& Makishima, K. 2013, \mnras,  427, 2824
\bibitem[Enoto et al.(2010a)]{Enoto10a}
Enoto, T., et al. 2010, \pasj, 62, 475 (Paper I)
\bibitem[Enoto et al.(2011)]{Enoto11}
Enoto, T. et al. 2011, \pasj,  63, 387
\bibitem[Gelfand \& Gaensler(2007)]{Gelfand+Gaensler}
Gelfand, J. \& Gaensler, B. M. 2007, \apj, 667, 1111
\bibitem[Haberl et al.(2006)]{Precession06}
Haberl, F.  et al. 2006, \aap, 451, L17
\bibitem[Harding \& Lai(2006)]{Harding+Lai06}
Harding, A. \& Lai, D. 2006, Rep. Prog.  Phys. 69, 2631
\bibitem[Heyl \& Hernquist(2002)] {Heyl02}
Heyl, J. S. \& Hernquist, L. 2002,  \apj, 567, 510
\bibitem[Ioka (2001)] {Ioka01}
Ioka, K. 2001, \mnras,  327, 639-662
\bibitem[Iokai \& Sasaki(2004)]{Ioka+Sasaki04}
Ioka, K. \& Sasaki, M. 2004,  \apj, 600, 296
\bibitem[Iwahashi \etal(2013)]{Iwahashi13} 
Iwahashi, T. et al. 2013, \pasj, 65, Art. 52
\bibitem[Kokubun \etal (2007)]{HXD2}Kokubun. M, et al.\ 2007c, \pasj, 59, S53
\bibitem[Koyama  \etal (1989)]{Koyama89}Koyama, K. et al.\ 1989, \pasj, 41, 469
\bibitem[Koyama  \etal (2007)]{XIS}Koyama, K. et al.\ 2007, \pasj, 59, S23
\bibitem[Kuiper et al.(2006)]{Kuiper06}
Kuiper, L.,  Hermsen, W.,  den Hartog, P. \& Collmar, W. 2006, \apj, 645, 556
\bibitem[Kuiper et al.(2012)]{Kuiper12}
Kuiper, L.,  Hermsen, W.,  den Hartog, P. R. \& Urama, J. O. 2012, \apj, 748, 133
\bibitem[Landau \& Lifshitz(1976)]{Landau+Lifshitz}
Landau, L. D.  \&  Lifshitz, E. M.  1976,
{\it Mechanics, Third Edition: Vol. 1 (Course of Theoretical Physics)},
Chapter IV (Butterworth-Heinemann).
\bibitem[Makishima(2014)]{Max14} 
Makishima, K.  2014,
in {The European Week of Astronomy
and Space Science (EWASS), Special Session 1} (June 30, Geneva)
\bibitem[Makishima et al.(2014)]{Makishima14} 
Makishima, K. et al. 2014, Phys. Rev. Lett. 112, id. 171102 (MEA14)
\bibitem[Mereghetti(2008)]{Mereghetti08}
Mereghetti, S. 2008, \aapr, 15,  225
\bibitem[Mitsuda \etal (2007)]{Suzaku}Mitsuda, K. et al.\ 2007, \pasj, 59, S1
\bibitem[Nakano et al.(2015)]{Nakano15}
Nakano, T. et al. 2015, \pasj, in press
\bibitem[Rea et al.(2012)]{Rea12}
Rea, N., et al. 2012, \apj, 754,  id. 27
\bibitem[Rea et al.(2013)]{Rea13}
Rea, N., et al. 2013, \apj, 770, id. 65
\bibitem[Rea et al.(2014)]{Rea14}
Rea, N., et al. 2014, \apjl, 781, id. L17
\bibitem[Ruderman \& Gil(2006)]{Ruderman06}
Ruderman, M \& Gil, J. 2006,  \aap, 460, L31
\bibitem[Shabanova et al.(2001)]{Precession01}
Shabanova, T. V., Lyne, A. G.,  \& Urama, J. O.  2001, \apj, 552, 231
\bibitem[Staubert et al.(2009)]{HerX1_09}
Staubert, R., et al. 2009,  \aap, 494, 1025
\bibitem[Takahashi \etal (2007)]{HXD}Takahashi, T., et al.\ 2007c, \pasj, 59, S35
\bibitem[Takiwaki \etal (2009)]{Takiwaki09}
Takiwaki, T., Kotake, K., \& Sato, K. 2009, \apj, 691, 1360
\bibitem[Terada \etal (2008)]{Terada08}
Tearda, T., et al. 2008, \pasj, 68, 387
\bibitem[Thompson \& Duncan(1995)]{Magnetar} 
Thompson, C.  \&   Duncan, R. C. 1995, \mnras {\bf 275}, 255
\bibitem[Tiengo et al.(2013)]{Tiengo13} 
Tiengo, A. et al. 2013, Nature 500, 312
\bibitem[Vink \& Kuiper(2006)]{Vink06}
Vink, J., \& Kuiper, L. 2006, \mnras, 370, L14
\end{thebibliography}
\end{document}